\documentclass[12pt]{iopart}
\usepackage{iopams}
\usepackage{hyperref}
\usepackage{color}
\usepackage{cite}
\usepackage{graphicx}
\usepackage{color}
\usepackage{caption}
\usepackage{subcaption}
\pdfoutput=1
 
\newcommand{\pic}[3]{
  \begin{figure}
    \centering
      \includegraphics[width=#2\textwidth]{#1}
      \caption{\label{img.#1} #3}
    \vspace{1ex}
  \end{figure}
}

\newcommand{\pictwosub}[3]{ % two pictures
\begin{figure}
\centering
\begin{subfigure}{.5\textwidth}
  \centering
  \includegraphics[width=\textwidth]{#1}
  \caption{}
\end{subfigure}%
\begin{subfigure}{.5\textwidth}
  \centering
  \includegraphics[width=\textwidth]{#2}
    \caption{}
\end{subfigure}
\caption{\label{img.#1} #3}
\end{figure}
}

\begin{document}

\title[]{Impact of ADC non-linearities on the sensitivity to sterile keV neutrinos with a KATRIN-like experiment}

\author{Kai Dolde$^a$, Susanne Mertens$^{a,b,c}$\footnote{corresponding author, smertens@lbl.gov }, David Radford$^d$, Tobias Bode$^c$, Anton Huber$^a$, Marc Korzeczek$^a$, Thierry Lasserre$^e$ and Martin Slezak$^c$}

\address{$^a$ Institute for Nuclear Physics, Karlsruhe Institute of Technology, Karlsruhe, Germany}
\address{$^b$ Nuclear Science Division, Lawrence Berkeley National Laboratory, Berkeley, USA}
\address{$^c$ Max Planck Institute for Physics, Munich, Germany}
\address{$^d$ Oak Ridge National Laboratory, Oak Ridge, USA}
\address{$^e$ Institut sur la recherche des loins fondamentales de l'Univers, CEA, Paris, France}

\begin{abstract}
ADC non-linearities are a major systematic effect in the search for keV-scale sterile neutrinos with tritium $\beta$-decay experiments like KATRIN. They can significantly distort the spectral shape and thereby obscure the tiny kink-like signature of a sterile neutrino. In this work we demonstrate various mitigation techniques to reduce the impact of ADC non-linearities on the tritium $\beta$-decay spectrum to a level of $<$ ppm. The best results are achieved with a multi-pixel ($\geq10^4$ pixels) detector using full waveform digitization. In this case, active-to-sterile mixing angles of the order of $\sin^2 \theta = 10^{-7}$ would be accessible from the viewpoint of ADC non-linearities. With purely peak-sensing ADCs a comparable sensitivity could be reached with highly linear ADCs, sufficient non-linearity corrections or by increasing the number of pixels to $\geq 10^5$. 
\end{abstract}
\textbf{keywords: }{sterile neutrinos, ADC non-linearities, tritium $\beta$-decay, KATRIN}\\

\submitto{\NJP}
\maketitle

%\begin{document}
\section{Introduction}
Sterile neutrinos are a well-motivated extension of the Standard Model of particle physics and can provide an explanation for the non-zero mass of active neutrinos \cite{whitepaper,numsm}. They are right-handed $SU(2)$ x $U(1)$ singulets and unlike active neutrinos do not even take part in the weak interaction. Hence, their detection is only possible via their mixing with the active neutrinos. If neutrinos are Majorana particles, Majorana mass terms $M_R$ can be introduced for the right-handed neutrinos, which can have an arbitrary scale. \\ Sterile neutrinos with a mass in the keV range are highly motivated from a cosmological point of view. They are suitable candidates for both Warm and Cold Dark Matter, being in good agreement with small- to large-scale structure observations \cite{vega,missingdwarf1, missingdwarf2,cusp1, cusp2,tbtf}.
A possible hint for a sterile neutrino with mass $m_s \approx 7~\mathrm{keV}$ and $\sin^2 (2\theta) \approx 10^{-10}$ was seen in a stacked XMM Newton (X-ray observatory) spectrum of galaxy clusters, however the origin and interpretation of the decay line are still controversial \cite{whitepaper, bulbul,contro1, contro2}.\\
In tritium $\beta$-decay, a small admixture of sterile neutrinos would show up as a kink-like signature in the continuous $\beta$-spectrum. This paper investigates the influence of Analog-to-Digital Converter (ADC) non-linearities on the sensitivity to sterile neutrinos with KATRIN-like $\beta$-decay experiments. Monte Carlo and analytical models, based on measured ADC non-linearities, were used to simulate the non-linearity-induced spectral distortions for tritium $\beta$-decay. In particular, we investigate different techniques of digitization, namely peak-sensing or waveform-digitization ADCs. Finally, we present novel means to suppress the effect of ADC non-linearities to a below a part-per-million level. The sensitivity of a KATRIN-like experiment to sterile neutrinos is calculated based on the considered scenarios of digitization.

\section{Tritium $\boldsymbol\beta$-decay and sterile neutrinos}
Tritium $\beta$-decay is a promising process by which to search sterile neutrinos. They would manifest themselves in a characteristic kink-like signature in the continuos $\beta$-decay spectrum. This section gives a short introduction to tritium $\beta$-decay and shows the potential of a KATRIN-like experiment with respect to a sterile neutrino search. 

\subsection{Tritium $\beta$-decay}
In tritium $\beta$-decay $^3_1 \mathrm{H}$ decays to $^3_2 \mathrm{He}$, an electron and an electron anti-neutrino.

\begin{equation}
^3_1 \mathrm{H} \rightarrow ^3_2 \mathrm{He}^+ + e^- + \overline{\nu}_e
\label{eq.tritiumdecay}
\end{equation}

The released decay energy is shared between the latter two, resulting in a continuous $\beta$-spectrum. The maximum kinetic energy of the electron is shifted to lower energies for a non-zero neutrino mass which makes single $\beta$-decay a suitable process for a direct neutrino mass measurement \cite{mainz, troitsk}. 

The tritium decay rate is given by

\begin{equation}
\frac{d\Gamma}{dE}=C\cdot F(E,Z=2)\cdot p \cdot (E + m_e) \cdot (E_0 -E)\sqrt{(E_0-E)^2-m_\nu^2},
\label{eq.tritium}
\end{equation}
where $E$ is the kinetic electron energy, $F$ is the Fermi function, $p$ the electron momentum, $m_e$ and $m_\nu$ are the electron and neutrino mass, and $E_0$ is the endpoint energy for $m_\nu = 0$. $C$ is a normalization factor given by 

\begin{equation}
\frac{G_F^2}{2\pi^3}cos^2\Theta_C|M|^2,
\end{equation}

where $G_F$ is the Fermi constant, $\Theta_C$ is the Cabbibo angle and $M$ denotes the nuclear transition matrix element.

\subsection{Sterile neutrinos with a KATRIN-like experiment}

The KArlsruhe TRItium Neutrino (KATRIN) experiment \cite{katrin_design} is designed to analyze the spectrum shape in a region close to the endpoint where the impact of the active neutrino mass is maximal. \\

The measured tritium $\beta$-decay spectrum with KATRIN is a superposition of spectra, corresponding to the individual mass eigenstates of the electron (anti-)neutrino. However, it is not possible to resolve the light mass eigenstates $m_i$ experimentally, therefore KATRIN is sensitive to the effective electron neutrino mass
\begin{equation}
m_{\nu_e}= \sqrt{\sum_i m_i^2 |U_{ei}|^2 },
\end{equation}
where $U_{ei}$ are entries of the PMNS matrix \cite{pmns_1, pmns_2}. 
A right-handed (sterile) neutrino is a flavour eigenstate and hence composed of mainly a new mass eigenstate, with a small percentage of the known light neutrino mass eigenstates.  Conversely, the electron neutrino would contain a small admixture of the new mass eigenstate $m_s$. In the case of keV-scale sterile neutrinos, there is a large mass splitting between the light neutrino mass eigenstates and $m_s$ which would result in a detectable superposition of tritium $\beta$-decay spectra if the mixing is strong enough. The resulting superimposed spectrum is given by

\begin{equation}
\frac{d\Gamma}{d\mathrm{E}}=\cos^2 \theta \left(\frac{d\Gamma}{d\mathrm{E}}\right)_{m_{\nu_e}} \Theta(E_0 -E -m_{\nu_e}) + \sin^2 \theta \left(\frac{d\Gamma}{d\mathrm{E}}\right)_{m_s} \Theta(E_0 -E -m_s),
\label{eq.superpos}
\end{equation}

where $\theta$ is the mixing angle between active and sterile neutrinos and $\Theta$ is the Heaviside step function. Figure \ref{img.figure1a} shows the impact of active-to-sterile mixing on the tritium $\beta$-decay spectrum ($m_s = 10~\mathrm{keV}$ and $\sin^2 \theta = 0.2$). It leads to a spectrum distortion and leaves a characteristic kink-signature in the spectrum.\footnote{Note that a realistic mixing angle could be of the order of $\sin^2 \theta < 10^{-6}$. The value $\sin^2 \theta = 0.2$ is solely used for didactic reasons to make the kink signature visible in the spectrum.} \\

\pictwosub{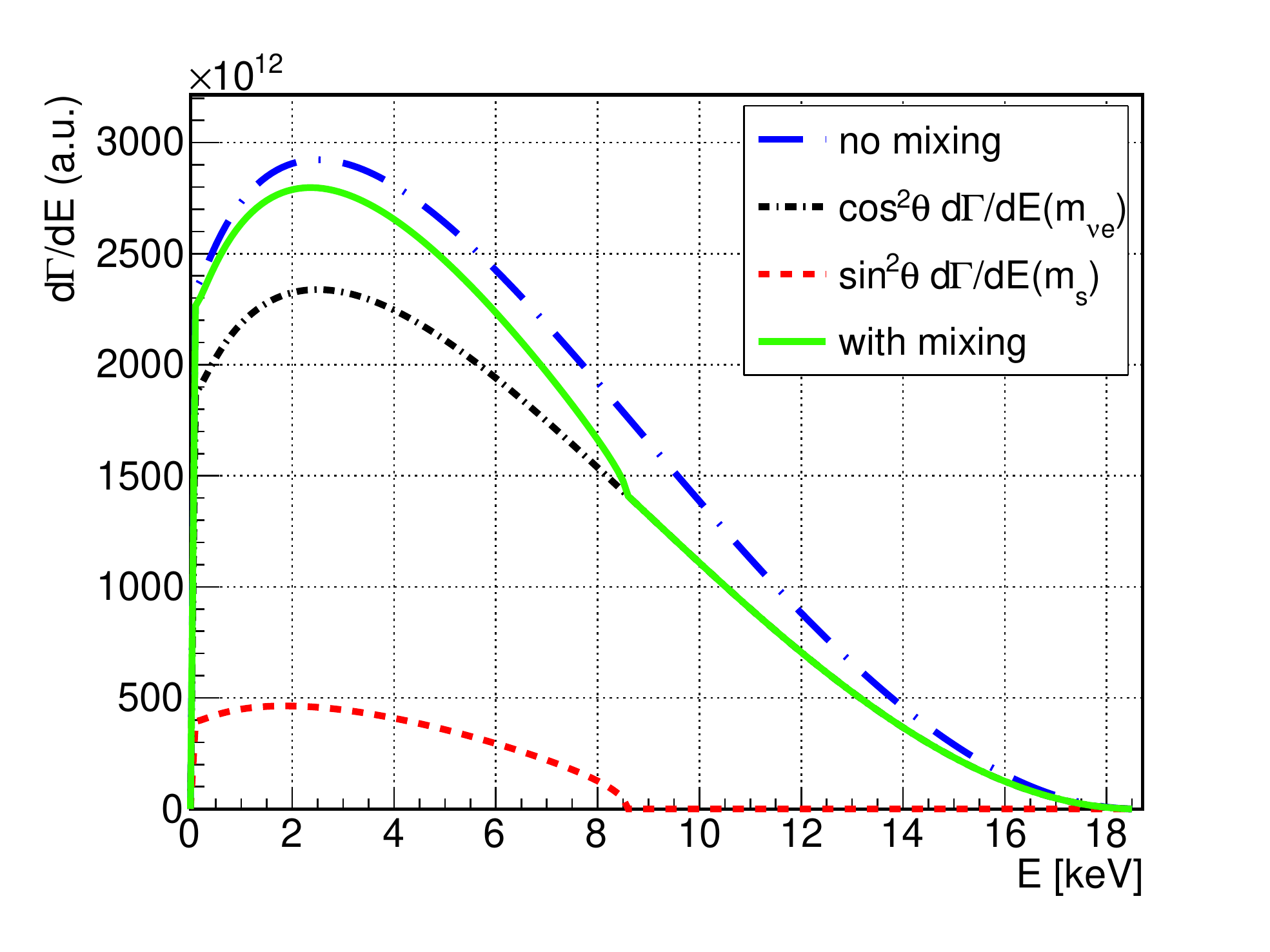}{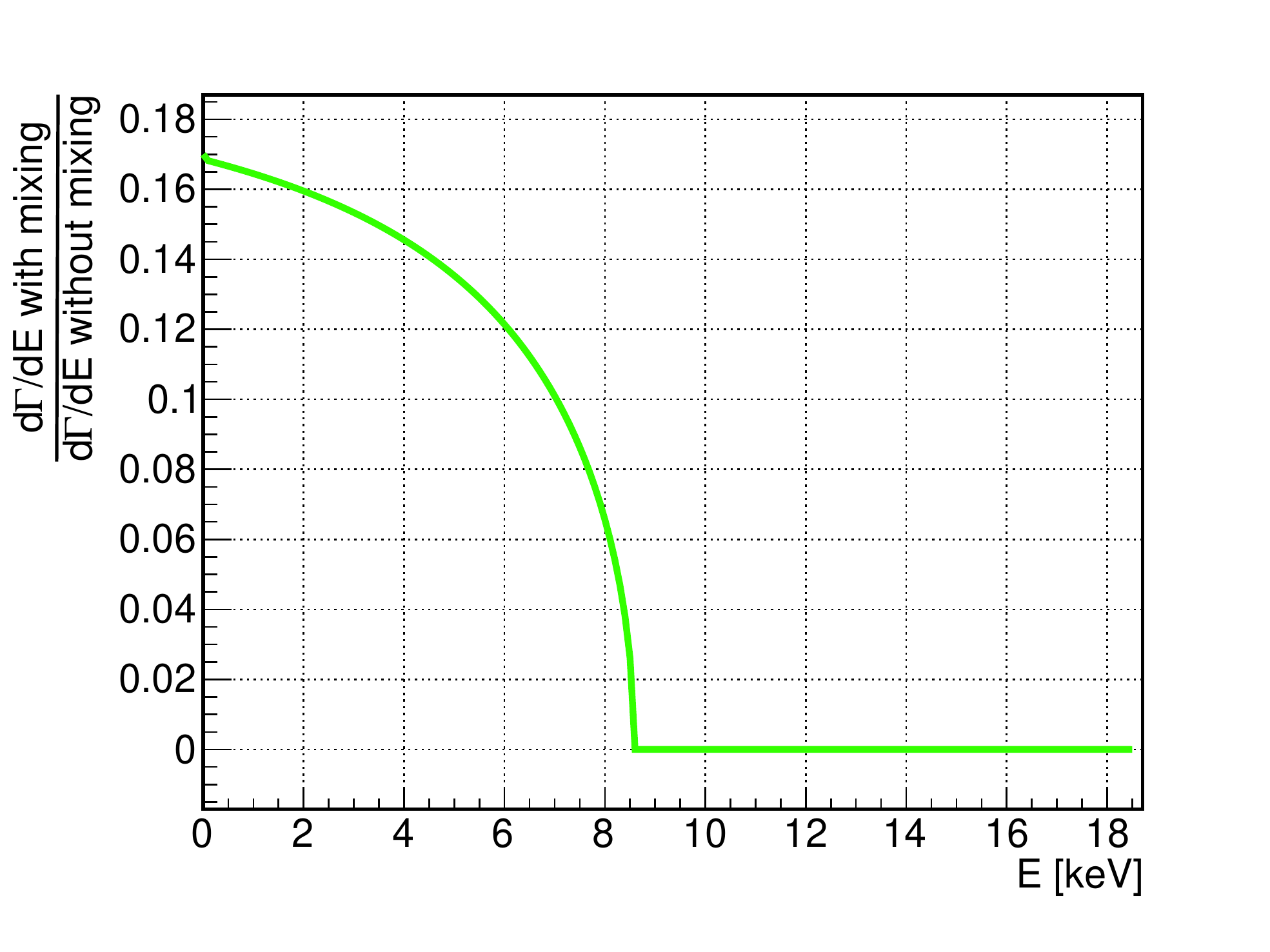}{\textbf{(a)} Tritium $\beta$-decay spectrum for $m_s = 10~\mathrm{keV}$ and $\sin^2 \theta = 0.2$ with active-to-sterile mixing (green solid line) and without mixing (blue dashed line). A characteristic kink-signature appears at the endpoint minus the heavy mass $m_\mathrm{s}$. \textbf{(b)} Ratio of spectra with and without mixing for $m_\mathrm{s}=10~\mathrm{keV}$ and $\sin^2 \theta = 0.2$. The ratio was shifted to zero above the kink.}

Recent studies \cite{wavelet, mertens_fit} show the high potential of a KATRIN-like experiment to search for sterile neutrinos in the keV range with a statistical sensitivity down to mixing angles of $\sin^2\theta \approx 10^{-8}$. The studies reveal that for a keV sterile neutrino search a differential measurement of the tritium $\beta$-decay spectrum is favoured over an integral measurement of the entire spectrum. In case of a differential measurement, one would no longer use the MAC-E-filter \cite{mace} technique to measure the energy (as done by KATRIN), but the detector itself would be used to measure the energy of each individual $\beta$ electron. This measurement mode, however, is prone to any non-linearities in the read-out system of the detector. \\
The unprecedented luminosity with $\approx 10^{11}~\mathrm{cps}$ of the windowless gaseous tritium source \cite{wgts} of KATRIN is advantageous for a keV-scale sterile neutrino search as it provides high statistics. Hence, it allows to probe small active-to-sterile mixing angles. However, with the nominal source strength of KATRIN, typical counting rates at the focal plane detector will be of the order of $10^{9}~\mathrm{cps}$. Since the current focal plane detector of KATRIN \cite{fpd} is not designed to handle these count rates, KATRIN would have to be equipped with a novel multi-pixel detector and read-out system for a high-statistics keV-scale sterile neutrino search.\\
With regard to this large-scale detector system, it is crucial to define the requirements on the read-out electronics. As shown in the following section, ADC non-linearties are a major systematic uncertainty in the sterile neutrino search in tritium $\beta$-decay with a future KATRIN-like experiment.

\section{ADC non-linearities}
\label{sec:ADCNonLinearities}
ADC non-linearities lead to errors on the measured particle energies and thus change the shape of the measured tritium $\beta$-decay spectrum. In the search for the small spectral distortion caused by a sterile neutrino, these modifications need to be understood and mitigated to a level smaller than the effect of the sterile neutrino we aim to detect. For example to be sensitive to a active-to-sterile mixing angle of the order of $<10^{-6}$ requires that ADC NL effects to be suppressed to less than 1 ppm, see section \ref{sec.sensitivity}. 
This section gives an introduction to ADC non-linearities, explains their origin and shows an example of a measured non-linearity spectrum of a Successive Approximation Register (SAR) ADC \cite{SAR, INL}. 

\subsection{Theory of ADC non-linearities}
\label{subsec:TheoryofADCNL}
Non-linearity is a very important parameter of every ADC and characterizes its imperfections and uncertainties due to the non-ideal components inside the ADC architecture. An ADC digitizes an input voltage with a certain resolution, depending on the number of bits which yields a quantization error. For an ideal ADC, the transfer function is a uniform step function with the same width for each step. However, due to the internal architecture with non-perfect components like capacitors and preamplifiers, as well as comparators and feedback Digital-to-Analog Converters, real ADCs show an intrinsic non-linear behaviour which cannot be eliminated by energy calibration. \\
This behaviour is described by the integral non-linearity (INL) which is defined as the deviation of the best-fit linear function to the ADC transfer function. Figure \ref{img.figure2} shows the schematic transfer and best-fit functions of an ADC. \\

Mathematically, the INL is defined as 

\begin{equation}
\mathrm{INL[}i\mathrm{]} = \frac{U[i]-U_\mathrm{0}}{U_{\mathrm{LSB}}} -i,
\end{equation}

where $i$ is the number of the respective step in the transfer function, $U[i]$ the corresponding input voltage, $U_\mathrm{0}$ the maximum input voltage to get 0 as ADC output code and $U_{\mathrm{LSB}}$ indicates the voltage which would cause an ideal ADC to increase its output by 1 Least Significant Bit (LSB). INL is typically given in units of LSB. 
  
\pic{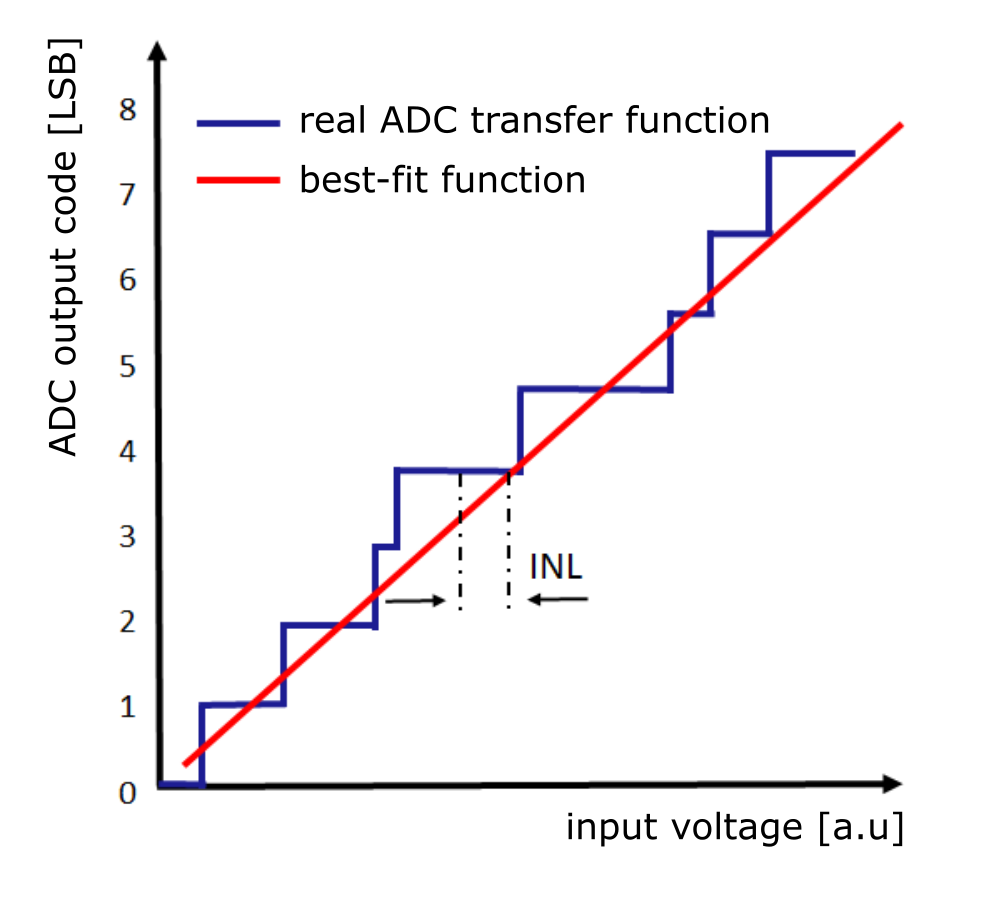}{0.4}{Schematic of the transfer function and linear best-fit function of a real ADC with a non-uniform step width. The deviation between the transfer and the best-fit function is called integral non-linearity.}

\subsection{INL of a SAR-ADC}
\label{subsec:INLsar}
For the purposes of illustration, the following sensitivity studies use example INL values, based on the measured INL spectrum of an \textit{Analog Devices} AD6645 SAR-ADC with a resolution of 14 bits, as it is used in the GRETINA experiment \cite{gretina}. A SAR-ADC compares the input voltage successively to comparator voltages of the DAC and becomes more precise with each step of comparison. Due to this self-repeating process, the same non-linearity structure appears consistently at different ADC output codes and shows a periodic pattern in the INL spectrum as shown in figure \ref{img.figure3}.

\pic{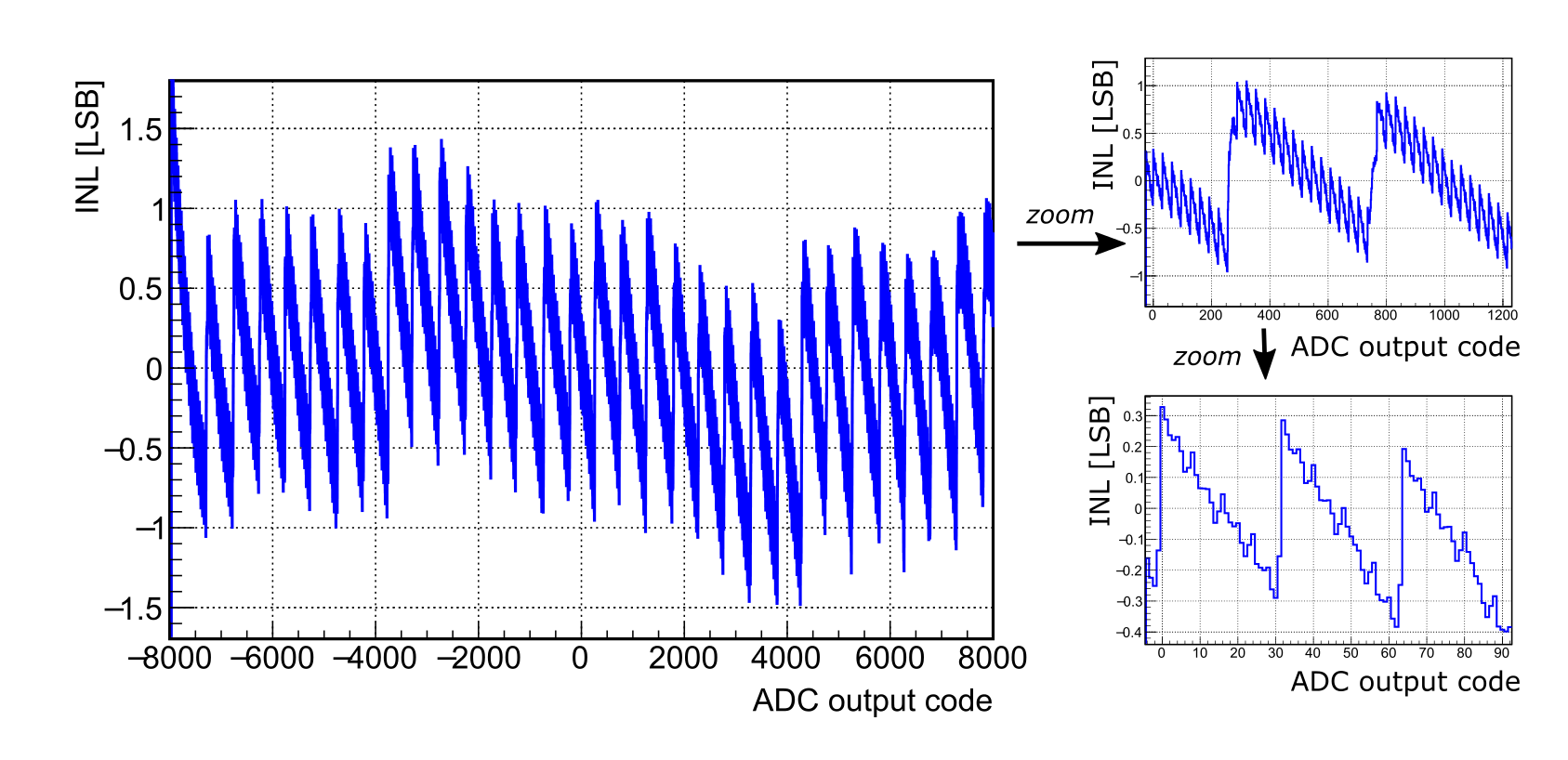}{1}{INL of an AD6645 (\emph{Analog Devices}) SAR-ADC of the GRETINA experiment \cite{gretina} with 14 bits. The periodic structure reveals the internal structure of the comparators inside the ADC as approximately a 5 bits/4 bits/5 bits comparator with three steps of successive approximations.}

\subsection{Model of INL}
\label{subsec:Model}
A model was developed to randomly generate INL spectra of a 14-bits SAR-ADC with three comparison steps of successive approximations in the ADC (5-bits, 4-bits, 5-bits), as it is used in the GRETINA digitizer as described above (section \ref{subsec:INLsar}).\\ The INL model is described by five parameters: gain, offset and three INL amplitudes according to the three comparison steps in the ADC. For each comparison step the INL amplitudes are drawn from a Gaussian distribution. The first comparison step leads to the large-scale structure in the INL spectrum, the subsequent ones generate the substructures. Additionally we allow for an offset that shifts the INL spectrum and a gain that can squeeze and stretch it. One example of a simulated INL spectrum is shown in figure \ref{img.figure4}.

\pic{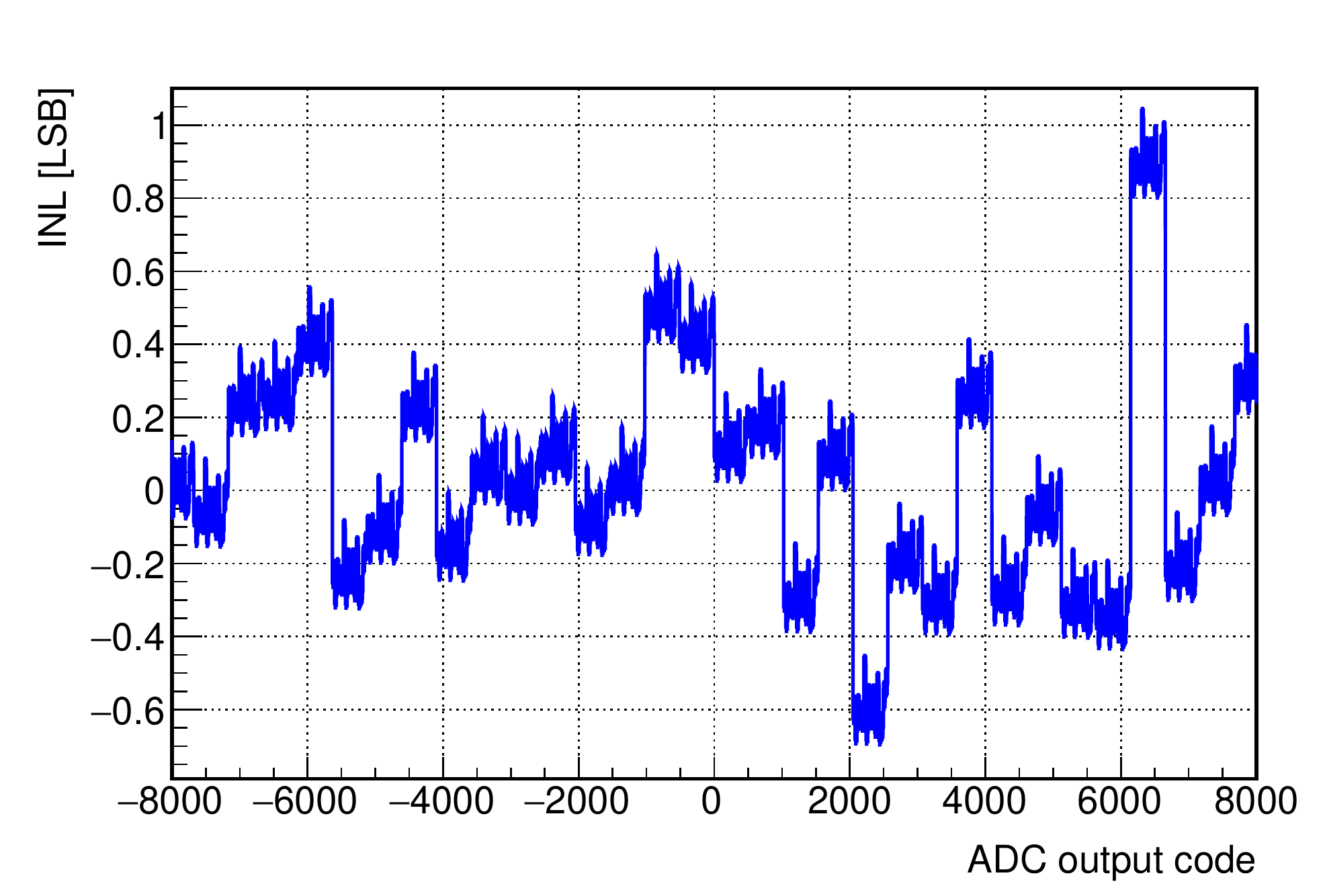}{0.8}{Simulated INL spectrum of a 14-bits SAR-ADC with three successive approximation comparison steps (5-bits, 4-bits, 5-bits).}

\section{Modelling of INL impact on tritium $\boldsymbol\beta$-decay spectrum}
\label{sec.modelling}
Having scrutinized the origin and properties of ADC non-linearities in section \ref{sec:ADCNonLinearities}, this section is dedicated to the impact of ADC non-linearities on measurements of the tritium $\beta$-decay spectrum. The tritium $\beta$-decay spectrum, including INLs, is modeled by simulating the realistic signals as created by a Si-detector and read-out system, assuming an illumination of the detector with tritium $\beta$-decay electrons at high rates. The simulations use input of the software $siggen$\cite{siggen}, which is a dedicated software to generate signals in semi-conductor detectors, used by several experiments such as GRETINA and MAJORANA~\cite{majorana}.

\subsection{Detailed description}
A pulse train, i.e. a long series of pulses, is generated whose energies are drawn from a probability density function, which corresponds to the tritium decay rate, given by equation \ref{eq.tritium}. The time between two pulses is drawn from an exponential distribution at a rate of $100~\mathrm{kHz}$. Figure \ref{img.figure5a} shows an excerpt of the resulting output pulse train of the MC simulation, i.e$.$ a pulse train, such as a continuously digitizing ADC would see. 

\pictwosub{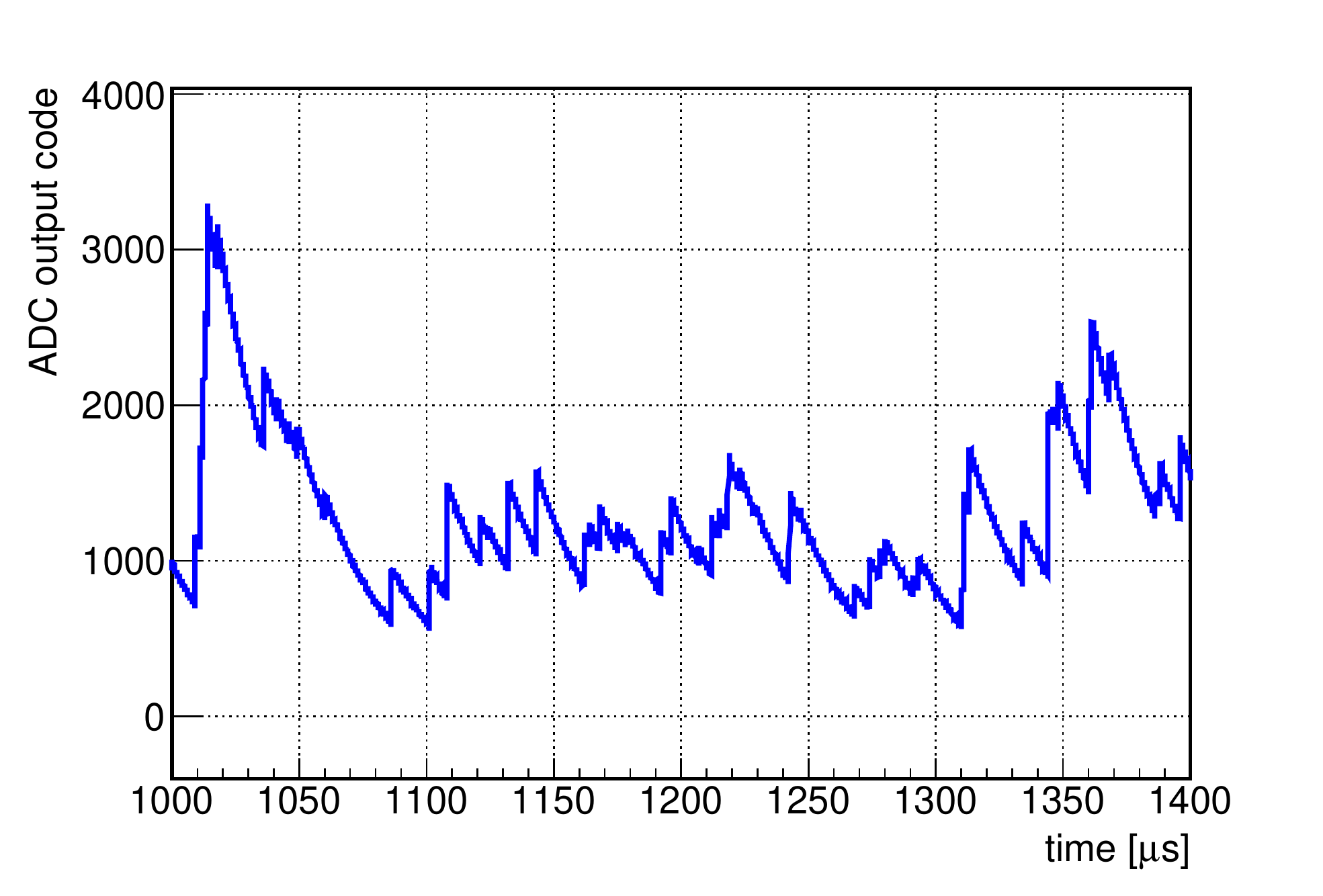}{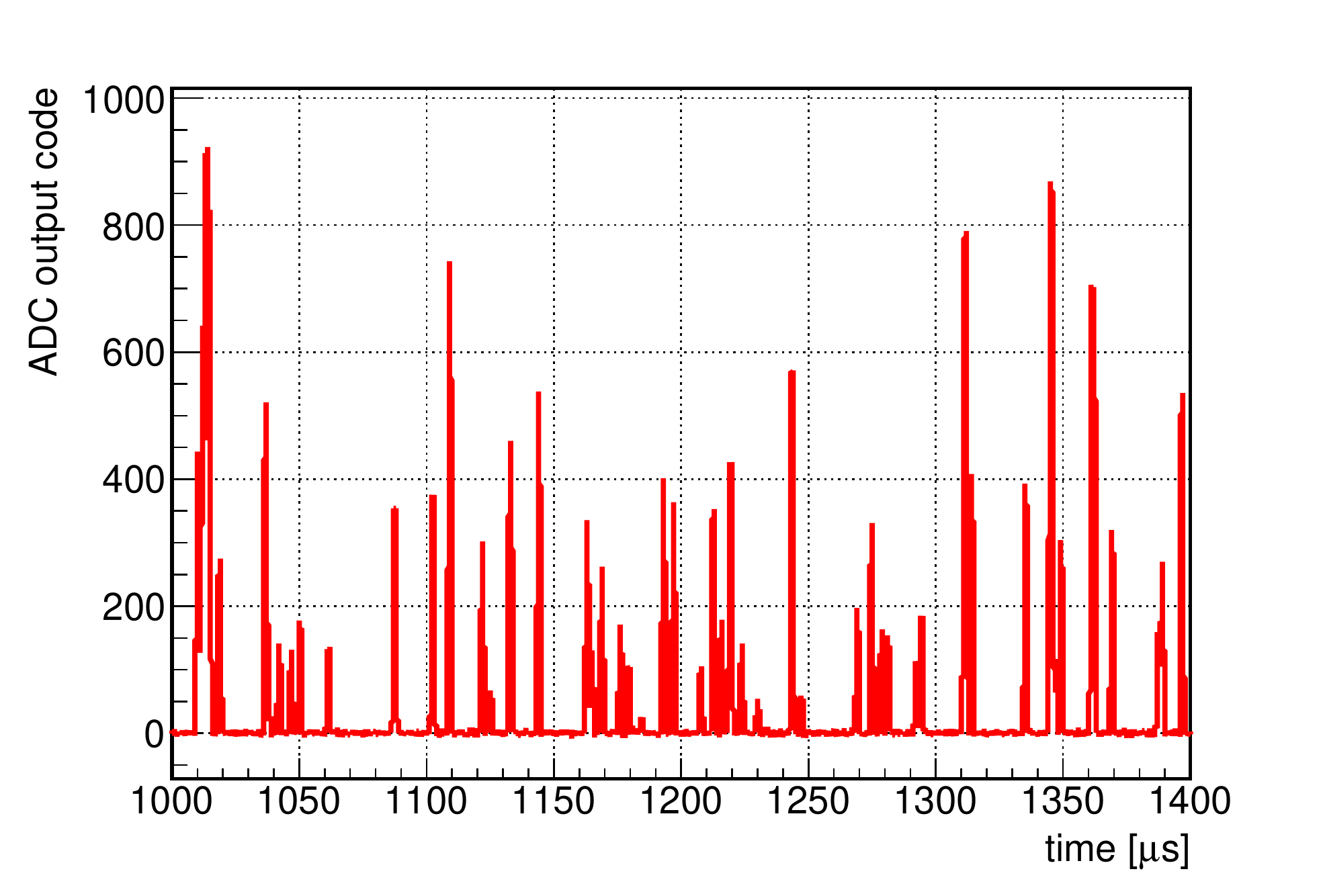}{\textbf{(a)} Excerpt of a simulated pulse train. Each pulse corresponds to an electron which creates a charge signal in a silicon detector. \textbf{(b)} After amplification and application of a trapezoidal filter, the peak heights of the shaped signals are used as a quantity which is proportional to the energies of the incident particles.}

The signal heights vary according to the particles' energies. Due to the high rate, the baseline does not return to ADC output code = 0, but the pulses often sit on the still high tail of a previous signal. This fact will be crucial when comparing the effect of non-linearities in the case of peak-sensing ADCs and waveform-digitizing ADCs. \\ 
To model the impact of ADC non-linearities the respective non-linearity values are added to the corresponding ADC output code of the waveform (figure \ref{img.figure5a}a) in case of a waveform digitizer, and to the ADC value of the energy (figure \ref{img.figure5a}b) in case of a peak-sensing ADC.\\
In our model we cover 1860 ADC output codes for electron energies up to 18.6 keV, due to pile-up up to $\approx 8000$ ADC output codes are covered, which corresponds to a dynamic range of $\approx 50\%$ of a 14-bits ADC. This allows for a shift of the entire energy spectrum by post-acceleration of the $\beta$-electrons as described in section \ref{subsec:pae}. The assumed INL amplitudes are in the order of $\pm 1 ~\mathrm{LSB}$.

\subsection{Simple case study}
Let us consider the simplest case of a 1-pixel detector and a peak-sensing ADC. To simulate the effect of non-linearities in this case only the output of a trapezoidal filter (figure \ref{img.figure5a}b) \cite{Trapezoidal_Grz, Trapezoidal_Jor} is digitized. A schematic working principle of a trapezoidal filter is shown in figure \ref{img.figure6a}.

\pictwosub{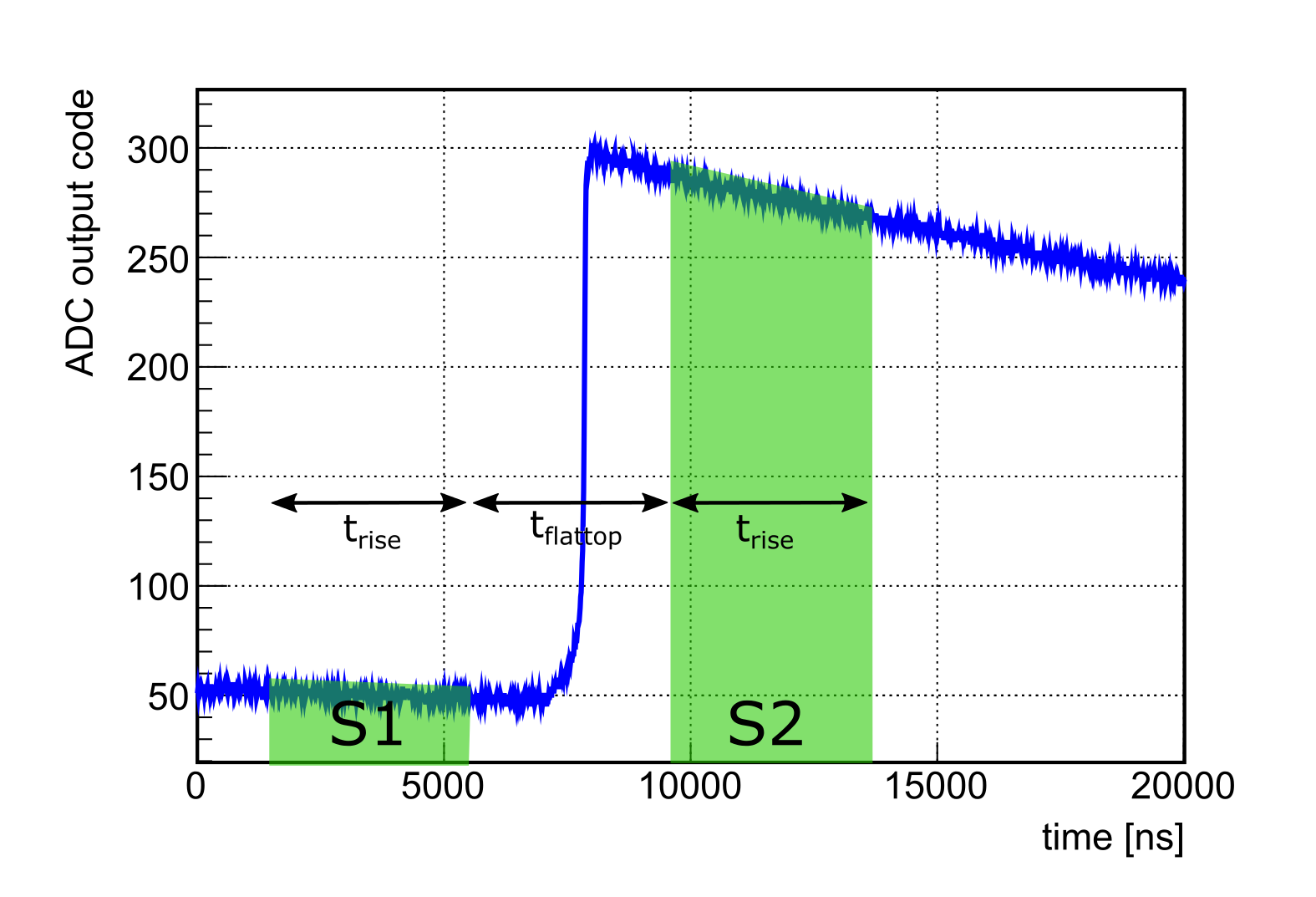}{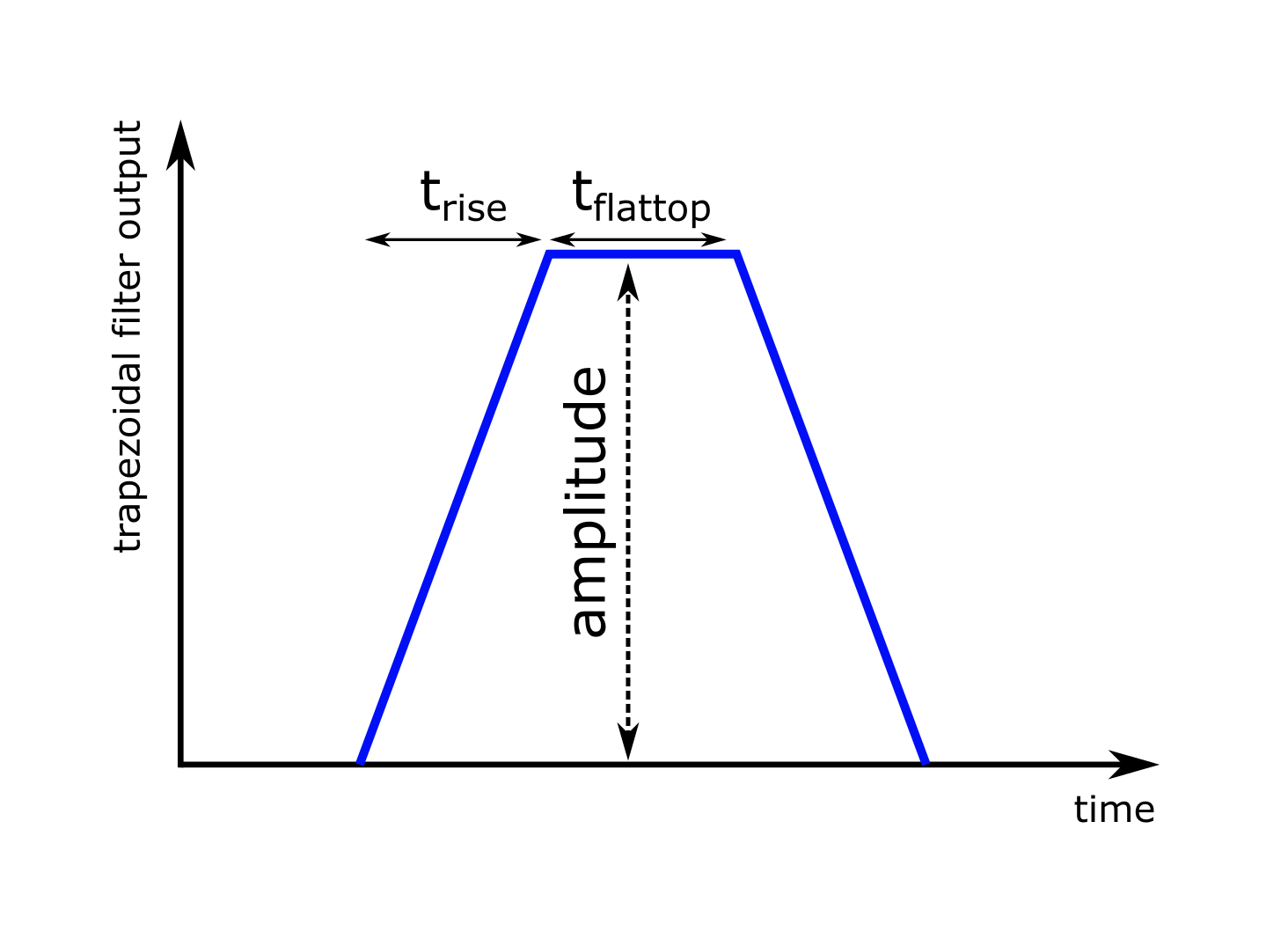}{Schematic of a trapezoidal filter. \textbf{(a)} For each point of the waveform the difference of the sums in the regions S2 and S1 is calculated and normalized to the number of sampling points in each of the regions. \textbf{(b)} Schematic of a trapezoidal filter output. The middle flat-top value of the filter output yields an amplitude proportional to the particle's energy.}

A MC pulse train of $10^8$ pulses is created and the energy, i.e. the output of a trapezoidal filter, is calculated for each pulse. Each energy is represented by an ADC output code. Subsequently, the corresponding INL value, obtained from the measurement shown in figure \ref{img.figure3}, is added to the respective ADC output code. Figure \ref{img.figure7a} shows the resulting tritium $\beta$-decay spectrum as it would be measured by a 1-pixel detector, equipped with a peak-sensing ADC. 

\pictwosub{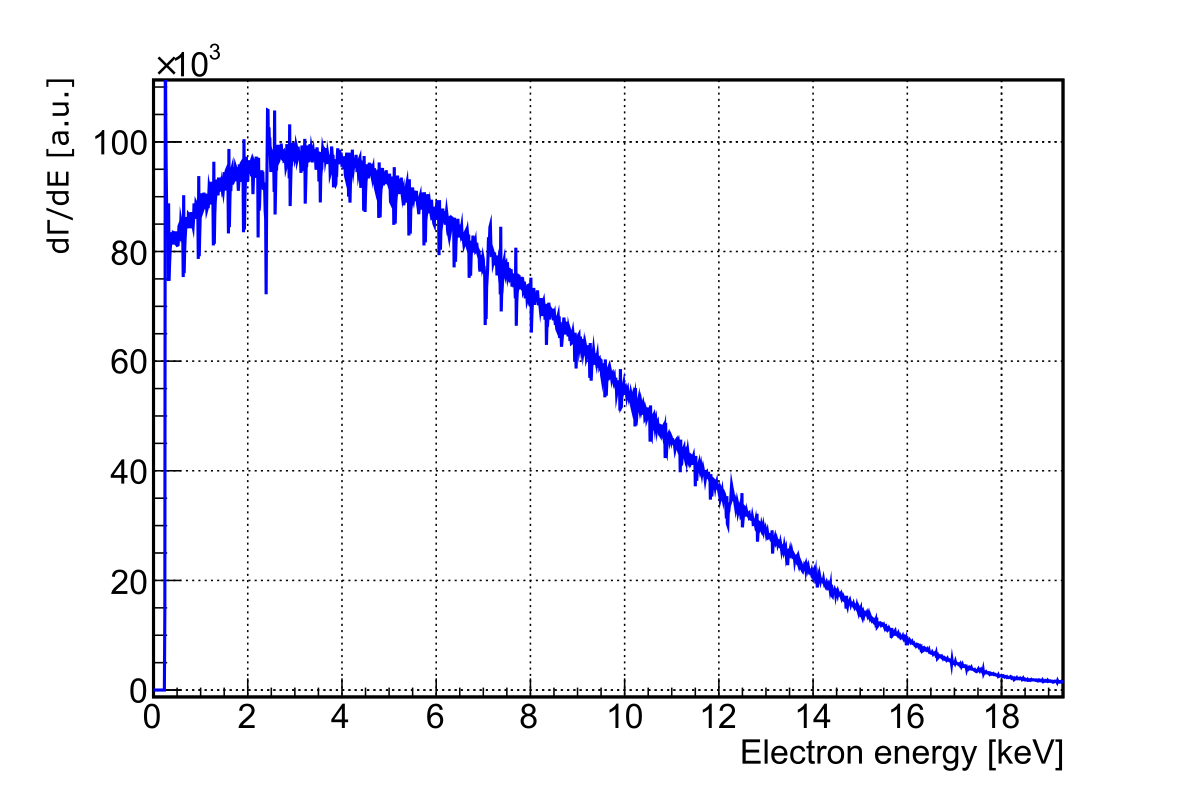}{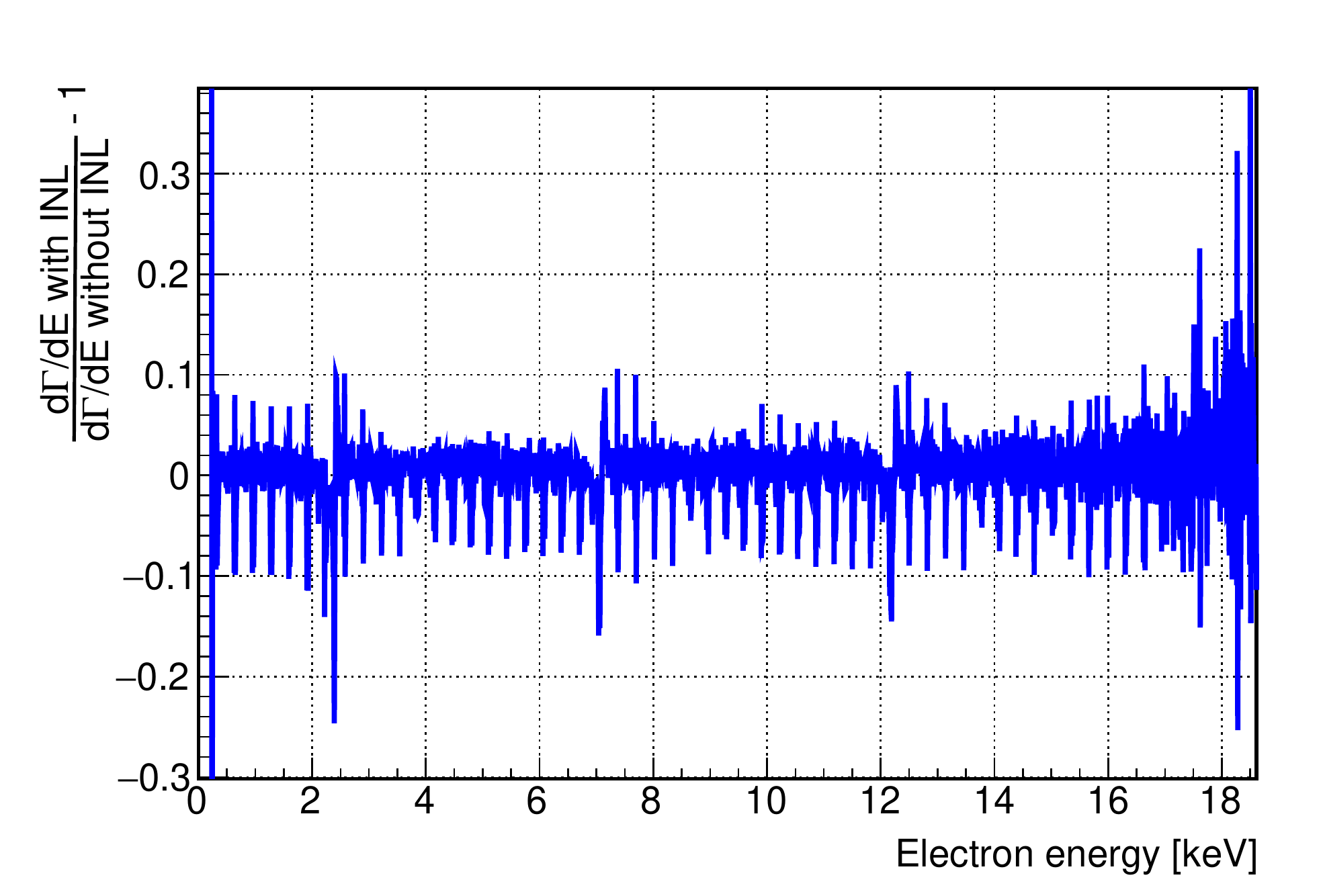}{\textbf{(a)} Tritium $\beta$-decay spectrum for $10^8$ simulated signal pulses, whose energy outputs from the trapezoidal filter are de-corrected according to the INL value of the respective ADC output codes, modelling the working principle of a peak-sensing ADC. The steps in the INL are directly transferred into the resulting energy spectrum. \textbf{(b)} Shifted ratio of tritium $\beta$-decay spectra with and without INL.}

The INL structure, as depicted in figure \ref{img.figure3}, directly translates into the energy spectrum and yields significant distortions of the spectrum shape. To quantify them, the ratio 
\begin{equation}
\frac{d\Gamma / dE ~\mathrm{with ~ INL}}{d\Gamma / dE ~\mathrm{without ~ INL}}-1
\end{equation}
is calculated. It shows INL-induced fluctuations of the order of $10 \%$ for an ADC coverage of 1860 out of $2^{14}$ ADC output codes. The large peaks correspond to the large-scale structure in the INL spectrum, caused by the first comparison step of the SAR-ADC.

\section{Means to mitigate ADC non-linearities}
In the simple case considered in section \ref{sec.modelling}, the resulting spectrum distortions widely exceed the expected distortions caused by a sterile neutrino kink-like signature, hence it is crucial to mitigate the INL-induced distortions.
In this section several approaches to mitigate the impact of ADC non-linearities are discussed. It is shown that usage of a \emph{Gatti}-slider, entire waveform digitization, additional post-acceleration and particular INL corrections can significantly smear out and reduce the INL-induced periodic structures.

\subsection{Gatti-slider}

To reduce the effect of the periodic structure in the INL, the pulses can be artificially shifted to different ADC ranges. This mitigation technique is known as a \emph{Gatti slider}, named after the Italian professor Emilio Gatti \cite{gatti}. A voltage is generated by a DAC of an N bit counter (e.g. N = 6) and is added to the analog input signal of the ADC. After digitization, the digital counter value is subtracted again to obtain the digital value of the pure input signal. The counter is increased by 1 after each pulse and is reset after a whole cycle of $2^N$ pulses. \\ Figure \ref{img.figure8a}a shows the ratio with/without INL for the usage of a 6-bit \emph{Gatti slider}. The spectrum distortions are reduced, but the bumps at the position of the big INL steps, e.g. at $2~\mathrm{keV}$, can still be seen. 

\subsection{Waveform-digitizing ADC}

Another approach is to digitize the entire waveform. In this case, the trapezoidal filter would be applied to the digitized waveform data (unlike in the case of a peak-sensing ADC, where only the output of the trapezoidal filter is digitized). This method has a major advantage in comparison to the peak-sensing ADC: It is sensitive to the baseline of each signal. As a consequence, different ADC outputs (with different INLs) are used for the measurement of the same energy and consequently, the INLs average out. 
As shown in figure \ref{img.figure5a}a, the baseline shifts widely through the ADC range due to the high signal rate. In contrast, in the peak-sensing case only the difference between baseline and signal height is digitized, making the result independent of the baseline. The non-linearity averaging due to the waveform digitizing is similar to the case of a \textit{Gatti slider}. However, in the case of the waveform digitizer it is included automatically, and works slightly more efficient as can be seen in figure \ref{img.figure8a}. 

\pictwosub{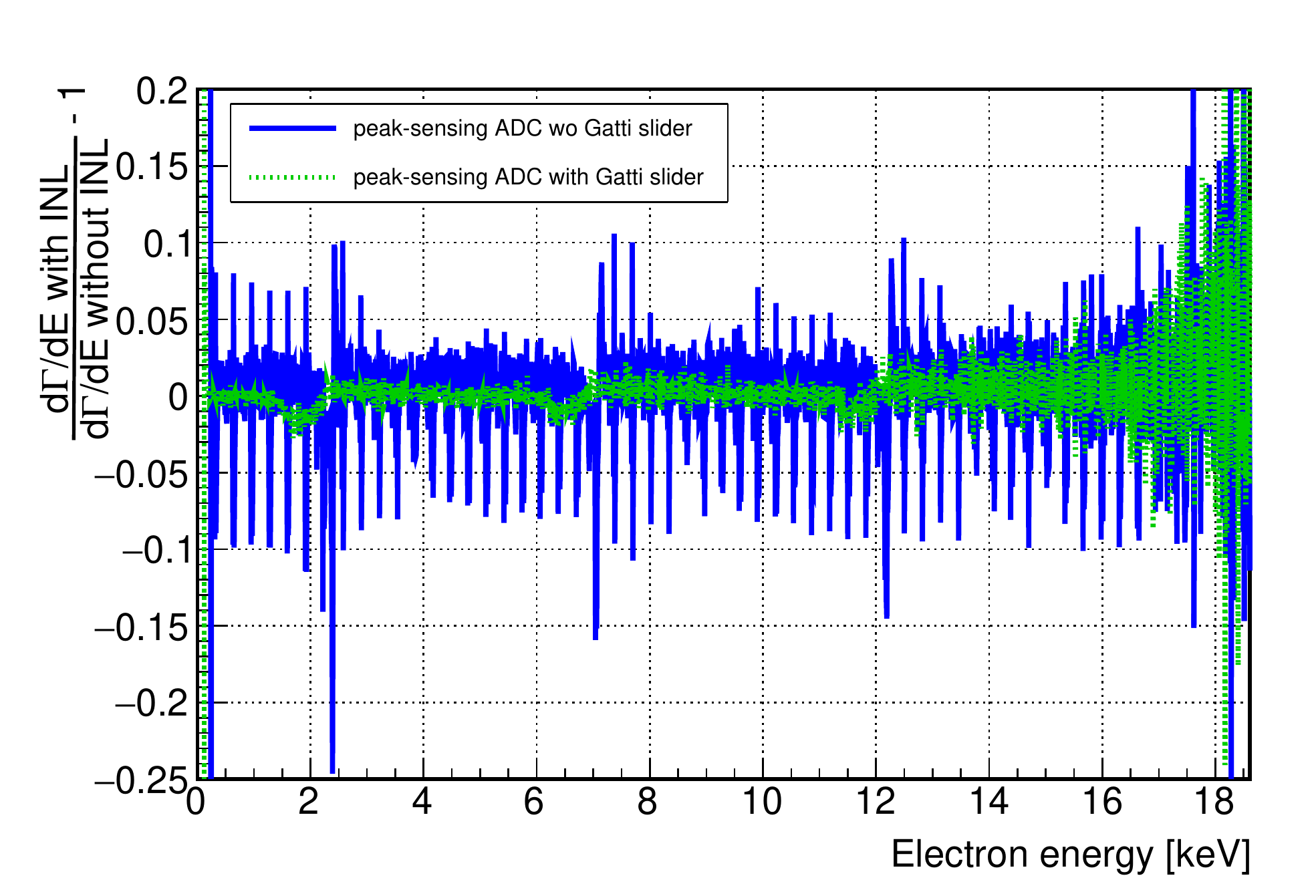}{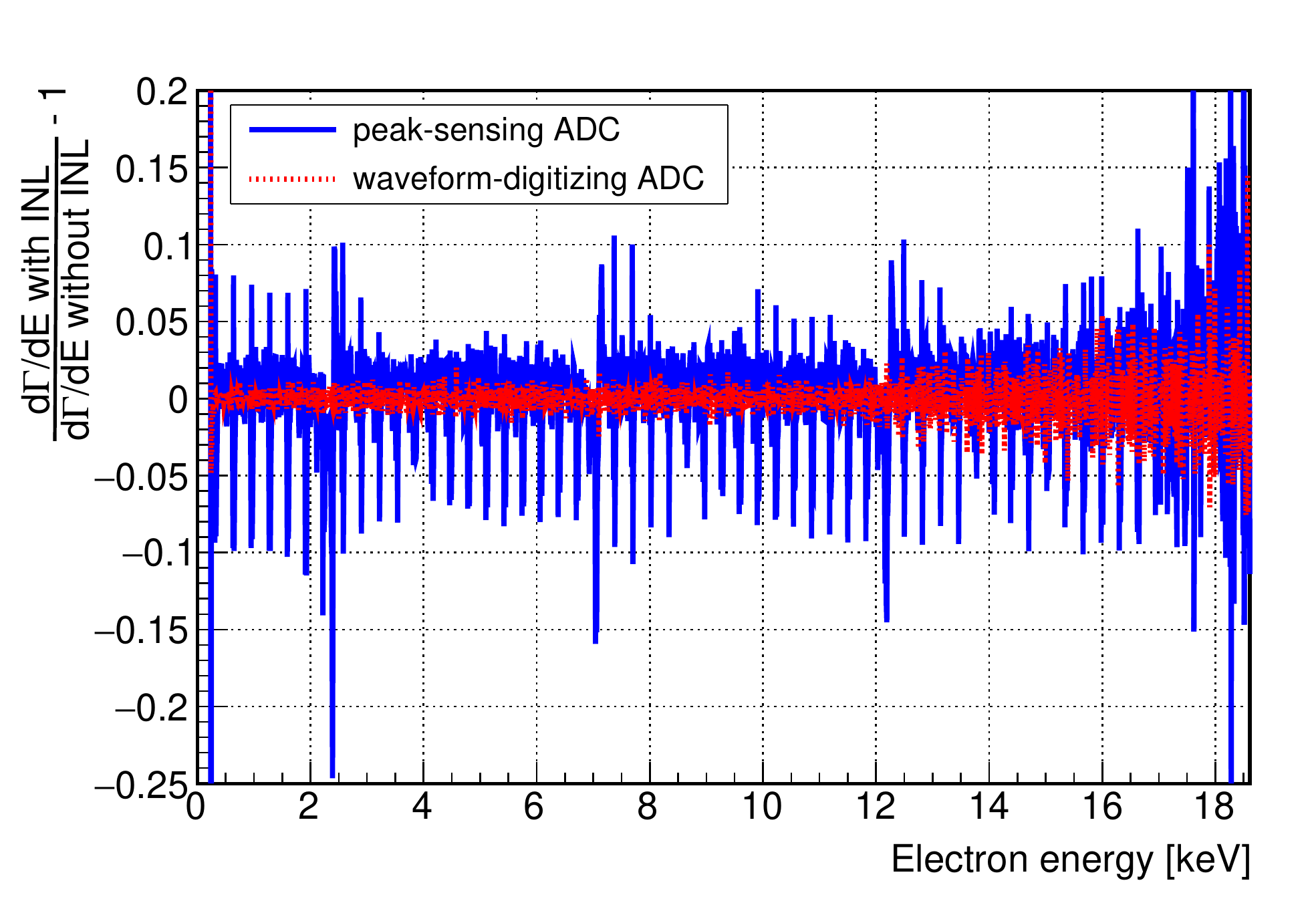}{\textbf{(a)} Ratio of tritium $\beta$-decay spectrum with and without INL, using a peak-sensing ADC. A \emph{Gatti slider} with a 6-bit counter significantly reduces the fluctuations in the ratio which occur because of the INL periodicity. The INL structure is partially averaged out. \textbf{(b)} Comparison of ratios for peak-sensing (no Gatti slider) and waveform-digitizing ADCs. For the waveform-digitizer the periodic INL structure is smeared out due to the baseline shifts and the resulting varying ADC ranges the waveforms cover.}

\subsection{Multi-pixel detector}
So far, we have only discussed a single detector pixel. However, for a sterile neutrino search in a KATRIN-like setup a multi-pixel detector with at least $10^4$ pixels is inevitable, in order to distribute the expected high count rate of $10^9$ cps (corresponding to the full KATRIN source strength) to many pixels, and thereby reduce the count rate per pixel. If each pixel is equipped with its own ADC, the effect of the INL is drastically reduced, since the INL of each pixel is slightly different and hence averages out.
To not be dominated by statistical fluctuations of a MC simulation, analytical simulations are required to simulate the impact of ADC non-linearities on the resulting tritium $\beta$-decay spectrum for a multi-pixel detector and a KATRIN-like counting rate of $10^9~\mathrm{cps}$. \\
Here we introduce the simulation models for both cases  of a peak-sensing ADC and waveform digitization and present the results. 

\subsubsection{Analytical simulation for peak-sensing ADCs} \hfill 
\label{subsec:AnapsADC}

The impact of ADC non-linearities on the tritium $\beta$-decay spectrum, recorded with a multi-pixel detector (here $N=10^4$ pixels), is investigated in the following way:

\begin{itemize}
\item A tritium $\beta$-decay spectrum without statistical fluctuations is calculated according to equation \ref{eq.tritium}.
\item Different INL spectra for each pixel$'$s ADC are generated, for each one the model parameters gain, offset and INL amplitudes are drawn from Gaussian distributions (see section \ref{subsec:Model}).
\item Applying an INL to an energy spectrum can be understood as streching or compressing energy bins. Accordingly, the INL is applied to the energy spectrum by redistributing a fraction of the bin contents to neighboring bins, depending on the respective INL value. \\
Demonstrating example: Assume that the calculated tritium $\beta$-decay spectrum has $x$ entries in the energy bin between the ADC output codes $y_3=3$ and $y_4=4$ with bin width $\delta y = y_4 - y_3 = 1$. Due to the applied INL the energies of the energy bins may change, so that for instance $y_{3,\mathrm{new}}=3.5$ and $y_{4,\mathrm{new}}=4.2$ with $\delta y_{\mathrm{new}}=0.7$. In this case $\frac{y_4-y_{3,\mathrm{new}}}{y_{\mathrm{new}}}=\frac{5}{7}$ of the $x$ entries remain in the new energy bin $y_{3,\mathrm{new}}$, whereas $\frac{y_{4,\mathrm{new}}-y_4}{y_{\mathrm{new}}}=\frac{2}{7}$ of the entries $x$ are shifted to the neighbouring bin $y_{4,\mathrm{new}}$. This procedure is applied to all energy bins. 
\item The N resulting energy spectra are summed up and normalized. 
\end{itemize}

\subsubsection{Analytical simulation for waveform-digitizing ADCs} \hfill \hfill

In the case of waveform-digitizing ADCs, the INL enters into the waveform of the signal, i.e. before the energy is calculated. How the INL eventually migrates to the reconstructed energy is shown in the following: \\

The energy output $E$ of a trapezoidal filter is calculated by

\begin{equation}
E = \frac{1}{t_{\mathrm{rise}}}\left(\sum_{\mathrm{S2}}\mathrm{wave}[t]-\sum_{\mathrm{S1}}\mathrm{wave}[t]\right),
\end{equation}

where the intervals S1 and S2 are defined in figure \ref{img.figure6a}. The values of $\mathrm{wave}[t]$ are digitized, and hence they are subject to INLs: 

\begin{equation}
\mathrm{wave}_{\mathrm{INL}}[t]  = \mathrm{wave}[t] + \mathrm{INL(wave[}t\mathrm{])},
\end{equation}

where $\mathrm{wave}_{\mathrm{INL}}[t]$ denotes the waveform including INL,  $\mathrm{wave}[t]$ the waveform without INL and $\mathrm{INL(wave[}t\mathrm{])}$ the respective INL value for the ADC output code at time step $t$. 

Consequently, the trapezoidal filter output, $E_\mathrm{INL}$ including INLs is given by

\footnotesize
\begin{eqnarray}
E_\mathrm{INL}& = &\frac{1}{t_{\mathrm{rise}}}\left(\sum_{\mathrm{S2}}\mathrm{wave}_\mathrm{INL}[t]-\sum_{\mathrm{S1}}\mathrm{wave}_\mathrm{INL}[t]\right) \nonumber \\
& = & \frac{1}{t_{\mathrm{rise}}} \left(\sum_{\mathrm{S2}}\left(\mathrm{wave}[t] + \mathrm{INL(wave[}t\mathrm{]})\right) -\sum_{\mathrm{S1}}\left(\mathrm{wave}_\mathrm{INL}[t]  - \mathrm{INL(wave}[t]\mathrm{)}\right)\right) \nonumber \\
& = & E + \frac{1}{t_{\mathrm{rise}}}\left(\sum_{\mathrm{S2}}\mathrm{INL(wave[}t\mathrm{]})-\sum_{\mathrm{S1}}\mathrm{INL(wave}[t]\mathrm{)}\right) 
\label{eq.trap_INL}
\end{eqnarray}
\normalsize

with $E$ being the trapezoidal filter output for the waveform without INL. 

In order to obtain $E_{\mathrm{INL}}$, we need to simulate $\mathrm{wave}[t]$ and the corresponding $\mathrm{INL(wave[}t\mathrm{])}$, see equation \ref{eq.trap_INL}. For a given energy, the values $\mathrm{wave}[t]$ depend on the baseline, as discussed before and shown in figure \ref{img.figure5a}a. Hence, for our model we need to take into account the different baseline values occurring in the measurement of the $\beta$-decay spectrum at high rates. This baseline distribution is obtained by making again use of the pulse-train simulation as illustrated in figure \ref{img.figure9a}. 

\pictwosub{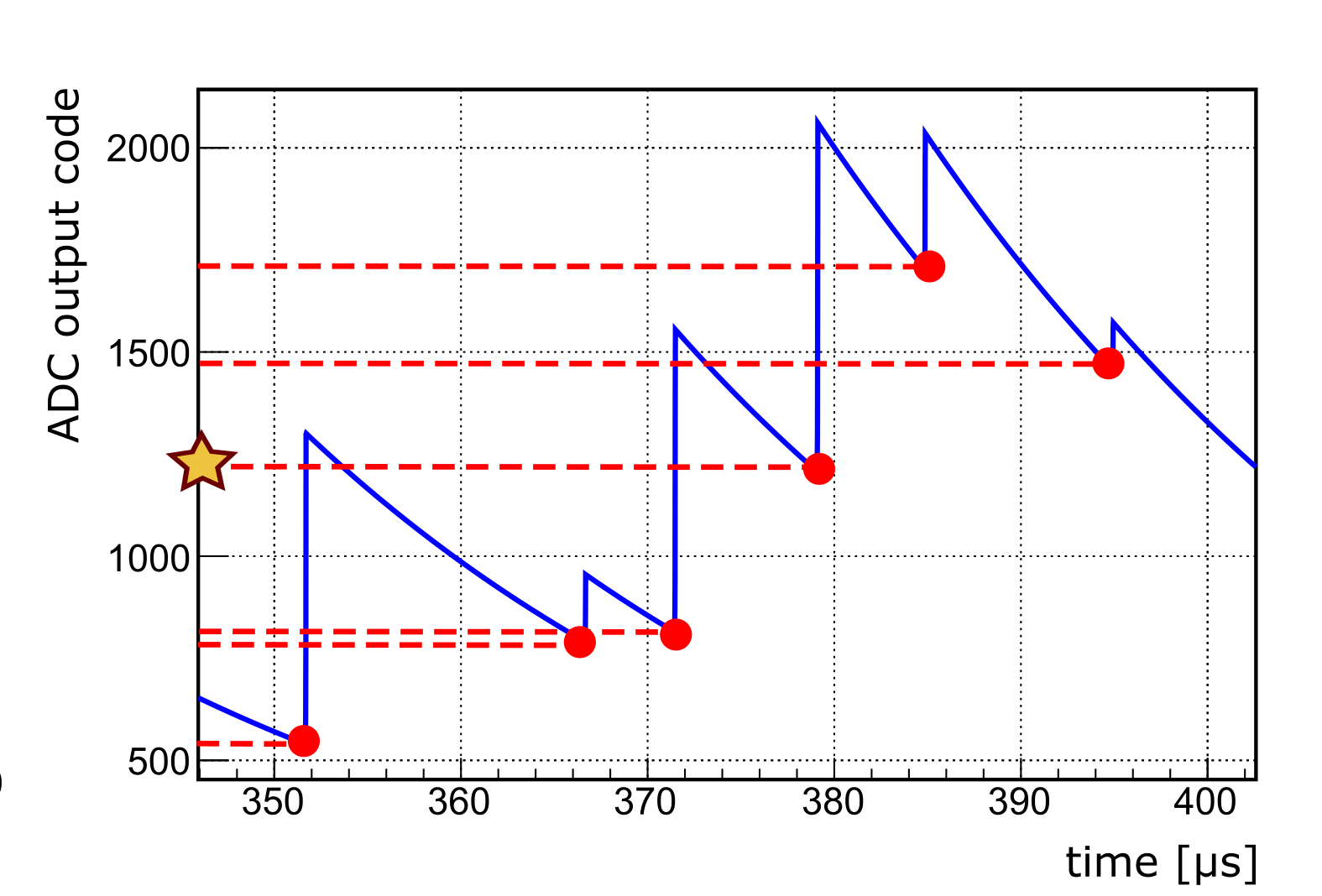}{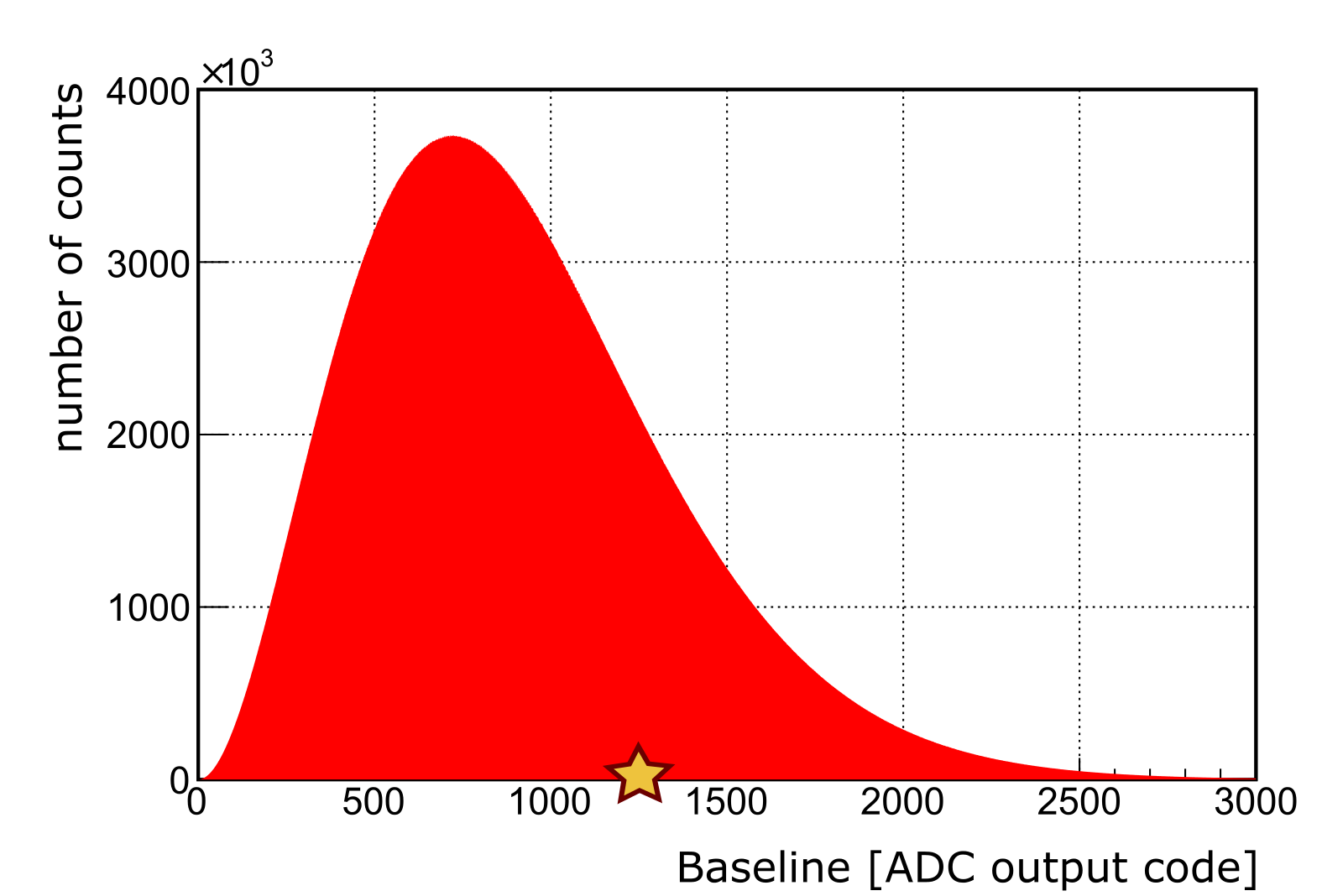}{\textbf{(a)} Simulated pulse-train. Due to the high rate the pulses sit on the tail of the previous pulse. Consequently the baseline of each signal varies. \textbf{(b)} The distribution of baselines as calculated from a pulse-train simulation shown in a) shows the frequency of the respective baseline ADC output codes of the simulated waveforms with the MC pulse train.}

For a given baseline and energy all values $\mathrm{wave}[t]$ can be easily calculated assuming a stable rise- and decay-time of the pulses. As in the case of peak-sensing ADCs, we slightly vary the INLs from pixel to pixel as described in section \ref{subsec:AnapsADC}. Putting these ingredients together we obtain the tritium $\beta$-decay spectrum, digitized by a waveform-digitizing ADC with the following procedure:
\begin{itemize}
\item The theoretically calculated tritium $\beta$-decay spectrum is convoluted with the baseline PDF. 
\item For each combination of baseline value $B$ and electron energy $E$, the trapezoidal filter output $E_\mathrm{INL}$ is calculated according to equation \ref{eq.trap_INL}.
\item The energy bin entries are fractionally redistributed to neighbouring bins, depending on whether $E_\mathrm{INL}>E$ or $E_\mathrm{INL}<E$.
\item The procedure is repeated N times with N different INL input spectra to simulate an N-pixel detector. The energy spectra are then added and normalized.
\end{itemize}

Figure \ref{img.figure10a} shows the ratios of the resulting tritium $\beta$-decay spectrum for a multi-pixel detector with $10^4$ pixels with and without INL, for both of the described analytical simulations methods of a peak-sensing and a waveform-digitization case. The fluctuations are two orders of magnitude smaller for the waveform-digitizing ADCs ($10^{-6}$) than for the peak-sensing ADCs ($10^{-4}$).

\pictwosub{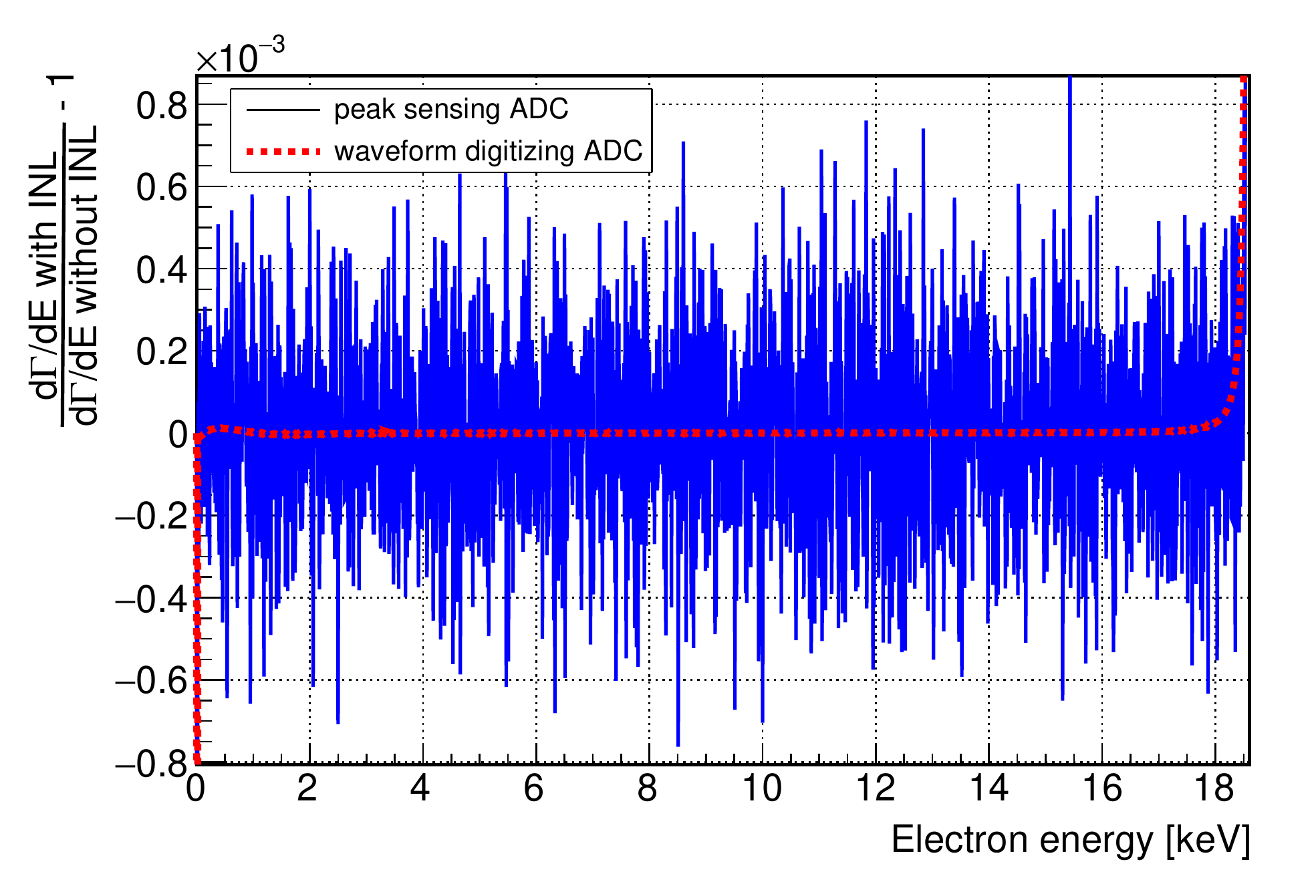}{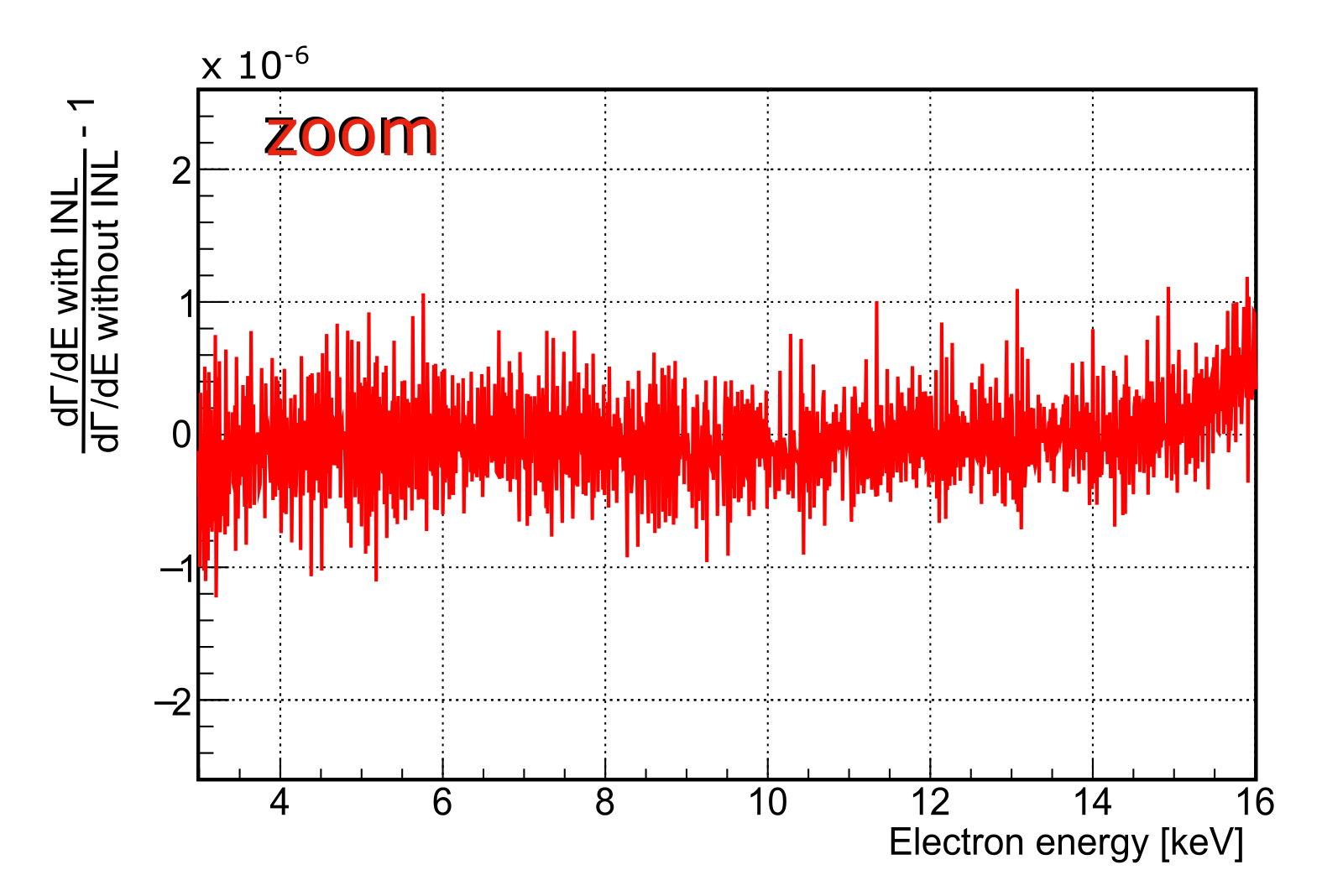}{\textbf{(a)} Ratio of tritium $\beta$-decay spectra with and without INL for the analytical simulation results for peak-sensing and waveform-digitizing ADCs for a multi-pixel detector with $10^4$ pixels. The fluctuations for the peak-sensing case are of the order of $10^{-4}$. \textbf{(b)} A zoom in the ratio for the waveform-digitizing case shows fluctuations of the order of $10^{-6}$.}

Divergent boundary effects occur, as the ratio of a spectrum with finite energy resolution (spectrum with INL) and a spectrum with infinite resolution (theoretical spectrum) is calculated. This effect is understood and its  correction will be the subject of follow-on work. 

\subsection{Mitigation by post-acceleration of electrons}
\label{subsec:pae}
The final possibility to mitigate the INL impact on the tritium $\beta$-decay spectrum discussed here, is to use a post-acceleration electrode which already exists in the current KATRIN setup. It can be used to shift the entire energy spectrum to higher energies by accelerating all electrons that pass the main spectrometer with an additional voltage, immediately before they hit the detector.

With post-acceleration, the ADC range of a waveform is not only shifted by a constant offset, but the particle energies themselves are changed, resulting in higher signals with steeper slopes. Signals with steeper slope cover more ADC output codes in the trapezoidal window S2 and the filter averages the INL values over a wider range of ADC codes.

Figure \ref{img.figure11a}a shows the original and five shifted tritium $\beta$-decay spectra, which are shifted by the post-acceleration electrode. In the data analysis, this additional post-acceleration energy would then be subtracted from the digitized energy. Figure \ref{img.figure11a}b displays the resulting ratio for a multi-pixel detector with waveform-digitizing ADCs and 100 steps of post-acceleration. The fluctuations are mitigated down to the order of $10^{-7}$.

\pictwosub{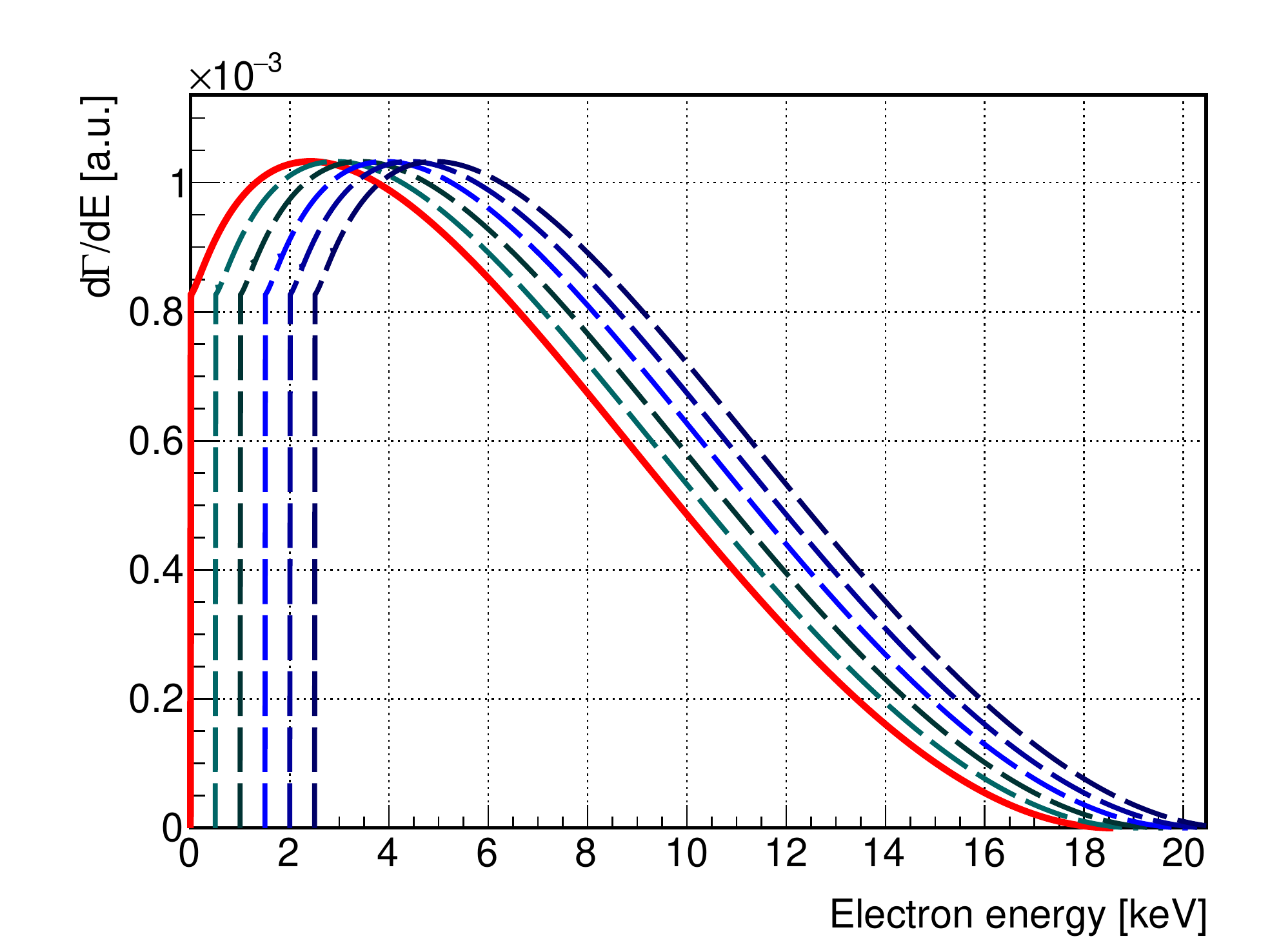}{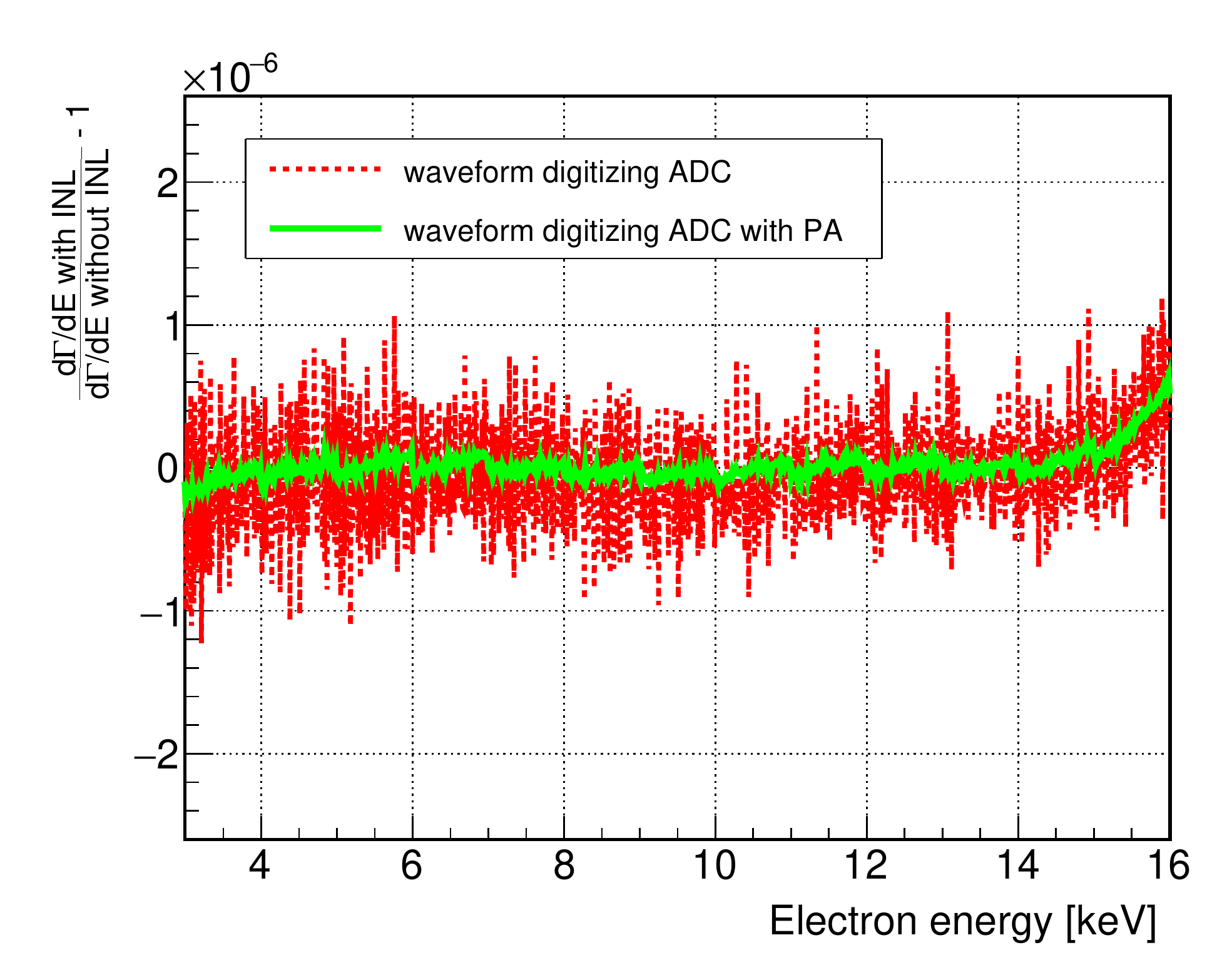}{\textbf{(a)} The original tritium $\beta$-decay spectrum (solid line) is shifted  by post-acceleration for different post-acceleration voltages (dashed lines). \textbf{(b)} Ratio of tritium $\beta$-decay spectrum with/without INL and addtional post-acceleration for a $10^4$ pixels detector with waveform-digitizing ADCs. Due to the post-acceleration the fluctuations are mitigated down to a level of $10^{-7}$.} 

\section{Sensitivity studies}
\label{sec.sensitivity}
In this section the impact of ADC non-linearities on the sensitivity in the sterile neutrino search with KATRIN is investigated. Covariance matrices \cite{cowan} are calculated for different digitization methods and different number of pixels, as discussed above. These matrices are then used as input for a statistical $\chi^2$-test to obtain $90\%$ exclusion limits. 

\subsection{Covariance matrices}
1000 KATRIN-like experiments were simulated, in which a specific number of pixels and a specific digitization method is assumed. In the 1000 experiments the INL inputs are varied, resulting in slightly different tritium $\beta$-decay spectra. Based on these spectra the covariance matrix is calculated.

Figure \ref{img.figure12a}  displays the covariance matrix for the case of a $10^4$ pixels detector with peak-sensing ADCs and additional post-acceleration of the electrons. The diagonal elements of the covariance matrix are much larger than the non-diagonal elements, indicating that the error is largely bin-to-bin uncorrelated and can, to a good approximation, be treated as equivalent to a statistical error. 
\pictwosub{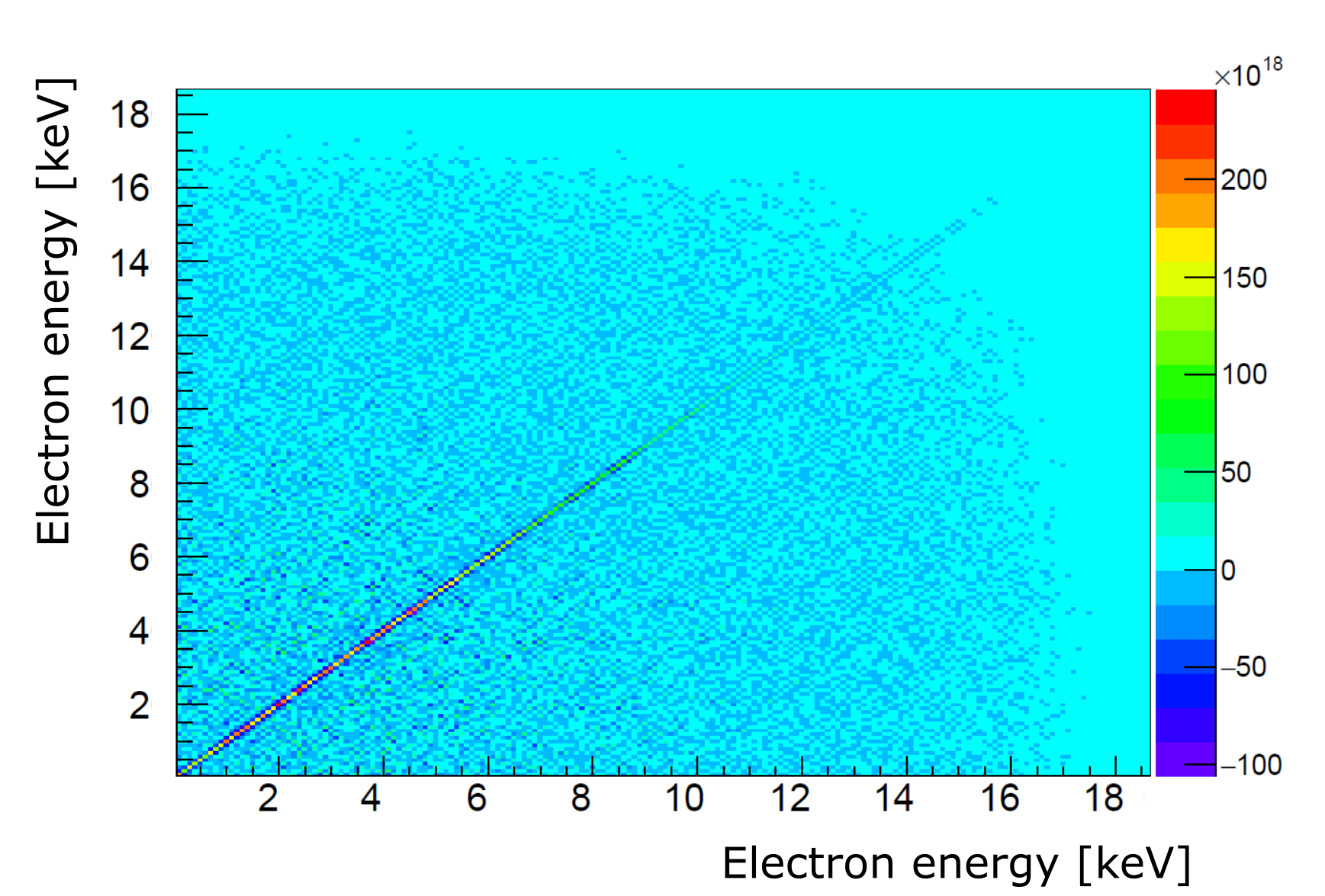}{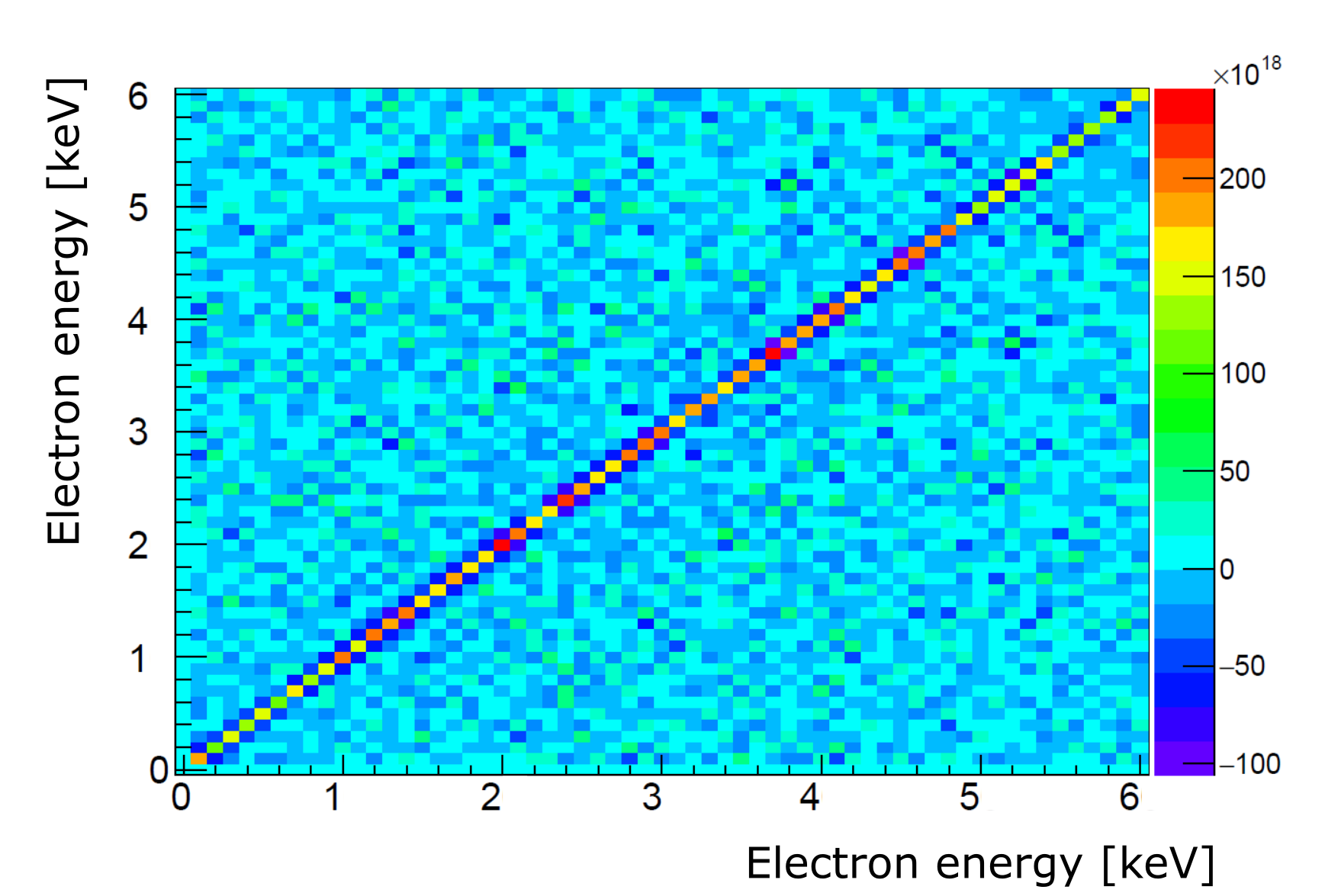}{Covariance Matrix of tritium decays of 1000 simulated KATRIN-like experiments with a $10^4$ pixels detector, read out by peak-sensing ADCs with additional post acceleration of the elctrons \textbf{(a)} and an expanded reqion of this matrix \textbf{(b)}. The INL fluctuations behave like an additional statistical error on the energy spectrum and do not insert significant correlations between the energy bins.}

\subsection{$\chi^2$-test}
Based on a $\chi^2$-test with $90\%$ confidence level the exclusion limits in the $m_s$-$\sin^2 \theta$-plane is calculated, see figure \ref{img.figure13}. 

\pic{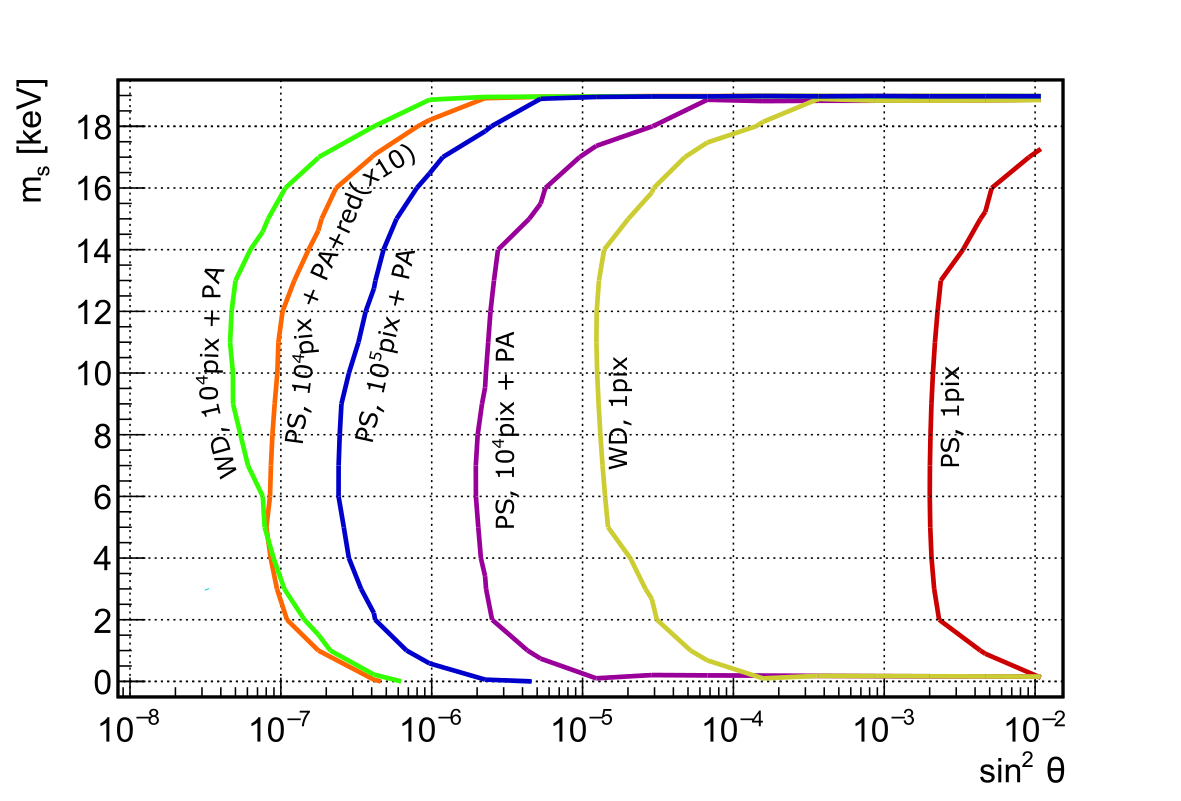}{0.8}{$90\%$ C.L. exclusion limit for a KATRIN-like experiment, assuming the full KATRIN source strength of  $9.4\cdot 10^{10}$ electrons per second and $3$ years of measurement time. The plot shows the KATRIN sensitivity with respect to a sterile neutrino  search for different detector and read-out configurations. For a detector with $10^4$ pixels and waveform-digitizing (WD) instead of peak-sensing (PS) ADCs with post-acceleration (PA) of the electrons, the impact of ADC non-linearities is mitigated and the sensitivity goes down to a mixing angle of $\sin^2 \theta = 10^{-7}$. Using peak-sensing ADCs sensitivity of  $\sin^2 \theta < 10^{-6}$ can either happen with an increased number of pixels from $10^4$ to $10^5$ or with an INL reduction by a factor of 10 by highly linear ADCs or an appropriate INL correction.}

For a one-pixel detector with a purely peak-sensing ADC only a sensitivity $\sin^2 \theta> 10^{-3}$ can be reached. Two orders of magnitude are gained by replacing the peak-sensing ADC with a waveform-digitizer. For a multi-pixel detector with $10^4$ pixels and additional usage of the post-acceleration electrode, a sensitivity down to $10^{-6}$ can be achieved with peak-sensing ADCs. Further improvement could be achieved  by reducing the INL for each ADC by a factor of ten, by for example using highly linear ADCs, or through INL measurements and corrections. By increasing the number of pixels to $10^5$ the sensitivity with peak-sensing ADCs can be further improved. Finally, the best-case sensitivity with respect to a sterile neutrino search and $10^4$ pixels is obtained with waveform-digitizing ADCs with additional post-acceleration. In this case a sensitivity of $\sin^2 \theta = 10^{-7}$ is obtained without any INL correction.

\section{Impact for final detector system}
In this section we relate our findings related to ADC non-linearities to a future KATRIN detector design. As mentioned before, with the full KATRIN tritium source strength a maximum signal rate of $10^{9}$ electrons/s can be achieved. With a minimal number of $10^4$ pixels the rate per pixel would be 100 kHz. We anticipate reducing the tritium source strength in the final experiment, to reduce systematic effects related to high-signal rates, such as pile-up.

From the point of view of ADC non-linearities two scenarios are conceivable: 
\begin{enumerate}
\item  a highly sophisticated ADC system with $10^4$ ADCs (with waveform digitization, or intrinsically low non-linearities), which in turn limits the number of pixels, due to costs and an complexity
\item a less sophisticated ADC system with larger number of pixels ($\geq 10^5$) and typical non-linearities \footnote{Generally, the number of ADCs is not fixed by the number of detector pixels. It is possible to read out a single pixel with multiple ADCs only for the purpose of reducing the effect of non-linearities.}
\end{enumerate}

Now, we would like to ask whether these requirements with respect to the ADC non-linearities are compatible with the general constraints on the pixel size and detector size: 
\begin{itemize}
\item The minimal pixels radius is determined by charge-sharing between neighboring pixels which makes the event reconstruction very challenging. To minimize charge-sharing, each pixel should have a radius $> 0.5~\mathrm{mm}$. In this case we expect to reduce the charge-sharing to less than $10\%$
\item The maximal detector radius is limited by the cyclotron radius of the electrons. For a large detector area $A$, the detector has to be placed in a low magnetic field $B$, so the magnetic flux $\Phi = B\cdot A$ is conserved in a KATRIN-like setup. In this case the cyclotron radius increases which would cause the electrons to hit surrounding structural material creating background, and would reduce the ability to veto backscattered electrons \cite{Marc}. Requesting a cyclotron radius $< 1~\mathrm{cm}$ sets a limit for the maximum detector radius of $35~\mathrm{cm}$. 
\end{itemize}

As can be seen in figure \ref{img.figure14} a detector with $10^4$ or $10^5$ pixels (as required by ADC non-linearities) perfectly matches the requirements with respect to pixel size and size of detector. These limits are not yet quantitatively fixed but provide an approximate order of magnitude. Further investigations will determine the exact parameters, taking into account also other criteria such as: pile-up, backscattering, energy loss in dead-layer, etc.

\pic{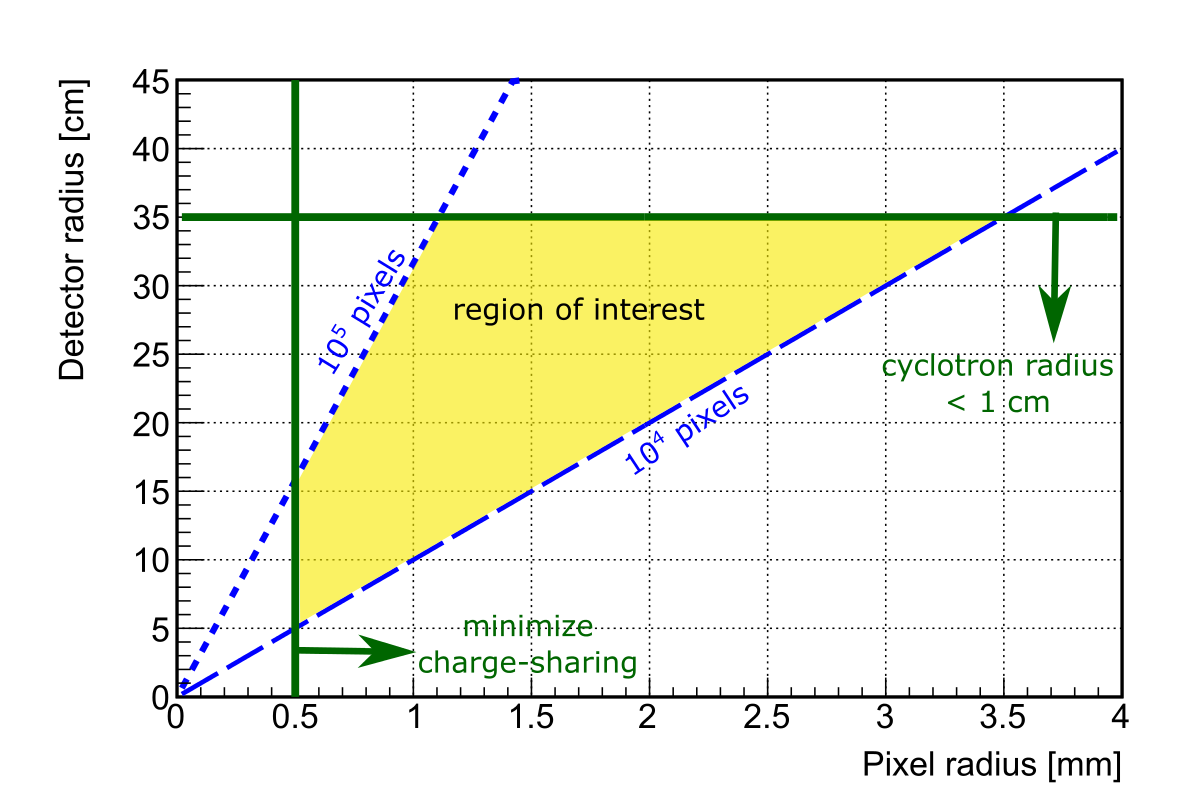}{0.8}{The total number of pixels, the maximum allowed cyclotron radius and the minimization of charge-sharing effects set limits on the pixel and detector radius and define a region of interest.}

\section{Conclusion}
In this paper, the impact of ADC non-linearities on the sensitivity of a KATRIN-like experiment to keV-scale sterile neutrinos is investigated. The model used here, is based on the INL of a 14-bits Successive Approximation Register ADC. We show that that a differential measurement with an energy-resolving detector is prone to ADC non-linearities, leading to significant distortions in the tritium $\beta$-decay spectrum. \\ We present a number of techniques to mitigate the effect of ADC non-linearities. Most notably, we demonstrate that the usage of a highly pixelated detector with at least $10^4$ pixels with waveform-digitizing ADCs and post-acceleration of the electrons drastically reduces the spectrum distortions and provides a sensitivity to sterile neutrinos down to a mixing angle of $\sin^2 \theta = 10^{-7}$ for a KATRIN-like experiment. With peak-sensing ADCs instead of waveform digitization, either the number of pixels has to be increased to $\geq 10^5$ or the used peak-sensing ADCs are required to show intrinsically ultra-low non-linearities to reach a sensitivity $<$ ppm.

\section*{Acknowledgements}
This work was supported by the German BMBF (05A14VK2) and by the Ministry of Science, Research and the Arts, Baden-W\"urttemberg (MWK). K. Dolde would like to thank KIT for financial support and both LBNL and ORNL for their hospitality.


\section*{References}
\begin{thebibliography}{99}

\bibitem{whitepaper}
R. Adhikari \textit{et al.}, \emph{A White Paper on keV Sterile Neutrino Dark Matter}, \emph{arXiv:1602.04816} (2016). 

\bibitem{numsm}
T. Asaka, M. Shaposhnikov, \emph{The nuMSM, dark matter and baryon asymmetry of the universe}, \emph{Physics Letters B} {\bf vol.620} (2005) 17-26.

\bibitem{vega}
H.J. de Vega and N. G. Sanchez, \emph{Cosmological evolution of warm dark matter fluctuations II: Solution from small to large scales and keV sterile neutrinos}, \emph{Phys. Rev. D} {\bf vol.85} (2012).

%\bibitem{wimp1}
%P. Gondolo, G. Gelmini, \textit{Cosmic abundances of stable particles: Improved analysis}, \textit{Nucl. Phys. B} {\bf vol.360} (1991) 145-179.

%\bibitem{wimp2}
%G. Jungman \textit{et al.}, \textit{Supersymmetric dark matter}, \textit{Phys. Rept.}, {\bf vol.267} (1996) 195-373

%\bibitem{wimp3}
%G. Gelmini, P. Gondolo \textit{Neutralino with the right old dark matter abundance in (almost) any supersymmetric model}, \textit{Phys. Rev. D}, {\bf vol.74} (2006) 

\bibitem{missingdwarf1} 
A. Schneider, \emph{Structure formation with suppressed small-scale perturbations},
\textit{Monthly Notices of the Royal Astronomical Society}, {\bf vol.451} (2015) 3117-3130.

\bibitem{missingdwarf2}
N. Menci \textit{et al.}, \emph{Galaxy Formation in WDM Cosmology}, \emph{Monthly Notices of the Royal Astronomical Society}, {\bf vol.421} (2012) 2384.

\bibitem{cusp1}
J. F. Navarro \textit{et al.}, \emph{A Universal density profile from hierarchical clustering}, \textit{Astrophysical Journal}, {\bf vol.490} (1997) 493-508.

\bibitem{cusp2}
David Weinberg \textit{et al.}, \textit{Cold dark matter: Controversies on small scales}, \textit{Proceedings of the National Academy of Sciences}, {\bf vol.112} (2015) 12249-12255.

\bibitem{tbtf}
E. Papastergis,  F. Shankar, \textit{An assessment of the "too big to fail" problem for field dwarf galaxies in view of baryonic feedback effects}, \textit{arXiv:1511.08741v1}, (2015)

\bibitem{bulbul}
Esra Bulbul \textit{et al.}, \textit{Detection of An Unidentified Emission Line in the Stacked X-ray spectrum of Galaxy Clusters}, \textit{Astrophysical Journal}, {\bf vol.789} (2014)

\bibitem{contro1}
O. Urban \textit{et al.}, \textit{A Suzaku Search for Dark Matter Emission Lines in the X-ray Brightest Galaxy Clusters}, \textit{arXiv:1411.0050}, (2014)

\bibitem{contro2}
Takayuki Tamura \textit{et al.}, \textit{An X-ray Spectroscopic Search for Dark Matter in the Perseus Cluster with Suzaku}, \textit{arXiv:1412.1869}, (2014)

\bibitem{mainz}
C. Kraus \textit{et al.}, \textit{Final results from phase II of the Mainz neutrino mass search in tritium beta decay}, \textit{European Physic Journal C}, {\bf vol.40} (2005) 447-468

\bibitem{troitsk}
V.M. Lobashev \textit{et al.}, \textit{Direct search for mass of neutrino and anomaly in the tritium beta-spectrum}, \textit{Physics Letters B}, {\bf vol.460} (1999) 227-235


\bibitem{katrin_design}
J. Angrik \textit{et al.}, \textit{KATRIN design report} (2004)


\bibitem{pmns_1}
B. Pontecorvo, \textit{Inverse beta processes and nonconservation of lepton charge}, \textit{Sov.Phys.JETP}, {\bf vol.7}, (1958) 172-173

\bibitem{pmns_2}
Z. Maki \textit{et al.}, \textit{Remarks on the unified model of elementary particles}, \textit{Progress of Theoretical Physics}, {\bf vol.28}, (1962), 870-880


\bibitem{wavelet} S. Mertens \emph{et al.}, \emph{Wavelet approach to search for sterile neutrinos in tritium $\beta$-decay spectra} \emph{Physical Review D}, {\bf vol.91} (2015).

\bibitem{mertens_fit} S. Mertens \emph{et al.}, \emph{Sensitivity of next-generation tritium beta-decay experiments for keV-scale sterile neutrino search}, \emph{Journal of Cosmology and Astroparticle Physics} {\bf vol.20} (2015).

\bibitem{mace} A.Picard \textit{et al.}, \textit{A solenoid retarding spectrometer with high resolution and transmission for keV electrons}, \textit{Nuclear Instruments and Methods in Physics B} {\bf vol.63}, (1992) 345 

\bibitem{wgts}
B. Bornschein, \textit{The closed Tritium cycle of KATRIN},\textit{Progress in Particle and Nuclear Physics}, {\bf vol.57}, (2006) 38-48

\bibitem{fpd}
J.F. Amsbaugh \textit{et al.}, \textit{Focal-plane detector system for the KATRIN experiment}, \textit{Nuclear Instruments and Methods in Physics Research A}, {\bf vol.778}, (2015) 40-60

\bibitem{SAR}
Linus Michaeli \textit{et al.}, \textit{Unified ADC nonlinearity error model for SAR ADC},\textit{ScienceDirect}, {\bf vol.41}, (2008) 198-204

\bibitem{INL}
Petr Suchanek \textit{et al.},\textit{ADC Nonlinearity Correction Based on INL(n) Approximations}, \textit{IEEE International Workshop on Intelligent Data Acquisition and Advanced Computing Systems: Technology and Applications},
(2009)

\bibitem{gretina}
I.Y. Lee \textit{et al.}, \textit{GRETINA: A gamma ray energy tracking array}, \textit{Nuclear Physics A}, {\bf vol.746}, (2004) 255-259

\bibitem{siggen}
Software developed by I-Yang Lee (LBNL), Karin Lagergren and David Radford (ORNL).

\bibitem{majorana}
C.E. Aalseth \textit{et al.}, \textit{The majorana $^{76}$Ge double-beta decay project}, \textit{Nuclear Physics B}, {\bf vol.124} (2003), 247-252

\bibitem{Trapezoidal_Grz}
Robert Grzywacz, \textit{Applications of digital pulse processing in nuclear spectroscopy}, \textit{Nuclear Instruments and Methods in Physics Research Section B}, (2003), {\bf vol.204}, 649-659.


\bibitem{Trapezoidal_Jor} 
V. T. Jordanov and G. F. Knoll, \textit{Digital synthesis of pulse shapes in real time for high resolution radiation spectroscopy}, \textit{Nuclear Instruments and Methods in Physics Research Section A}, (1994), {\bf vol.345}, 337-345.

\bibitem{gatti}
Emilio Gatti \textit{et al.}, \textit{A new method for analog to digital conversion}, \textit{Nuclear Instruments and Methods}, (1963), 241-242.

\bibitem{cowan}
G. Cowan, \textit{Statistical Data Analysis}, \textit{Oxford Science Publications}, (1997)

\bibitem{Marc}
M.Korzeczek \textit{et al.}, to be published


\end{thebibliography}
\end{document}